\documentclass[a4paper]{llncs}

\usepackage{epsfig}

\newtheorem{thm}{Theorem}
\newtheorem{lem}{Lemma}

\newtheorem{fig}{Figure}

\def\leurre{\noindent\leftskip0pt\small\baselineskip 10pt}

\def\encadre#1#2{%
\setbox100=\hbox{\kern#1{#2}\kern#1}
\dimen100=\ht100 \advance \dimen100 by #1
\dimen101=\dp100 \advance \dimen101 by #1
\setbox100=\hbox{\vrule height \dimen100 depth \dimen101\box100\vrule}
\setbox100=\vbox{\hrule\box100\hrule}
\advance \dimen100 by .4pt \ht100=\dimen100
\advance \dimen101 by .4pt \dp100=\dimen101
\box100
\relax
}

\def\ligne#1{\hbox to \hsize{#1}}
\def\PlacerEn#1 #2 #3 {\rlap{\kern#1\raise#2\hbox{#3}}}

\setbox221=\hbox{\includegraphics{cercle_L20.eps}}

\def\encercle#1#2{\hbox{\raise-5pt\copy221\hskip#2#1}}

\title{Constructing a uniform plane-filling path in the 
ternary heptagrid of the hyperbolic plane}
\author{Maurice Margenstern\inst{1}}
\institute{%
Laboratoire d'Informatique Th\'eorique et Appliqu\'ee, EA 3097,\\
Universit\'e de Metz, I.U.T. de Metz,\\
D\'epartement d'Informatique,\\
\^Ile du Saulcy,\\
57045 Metz Cedex, France,\\
\email{margens@univ-metz.fr}
}
\begin{document}
\maketitle

\vskip 10pt
\begin{abstract}
In this paper, we distinguish two levels for the plane-filling property.
We consider a simple and a strong one. In this paper, 
we give the construction which proves that the simple
plane-filling property also holds for the hyperbolic plane. 
The plane-filling property was established for the Euclidean plane by J. Kari, 
see~\cite{jkari94}, in the strong version. 
\end{abstract}
\vskip 10pt

\def\cqfd{\hbox{\kern 2pt\vrule height 6pt depth 2pt width 8pt\kern 1pt}}
\def\Hii{$I\!\!H^2$}
\def\Hiii{$I\!\!H^3$}
\def\Hiv{$I\!\!H^4$}
\def\norm{\hbox{$\vert\vert$}}
\section{Introduction}

   Consider a finite set of tiles~$T$ based on a regular polygon of the 
hyperbolic plane. We say that there is a {\bf solution} for tiling the
hyperbolic plane with tiles of~$T$, if and only if there is a 
partition~$\cal S$ of the hyperbolic plane such that the closure of each part
of~$\cal S$ is a copy of some tile of~$T$, where a copy of the figure~$F$ is
an isometric image of~$F$. We may adjoin conditions with
colours on the edges: we then require that adjacent tiles always define the
same colour on their common edge.

   The {\bf simple plane-filling property} consists in finding a finite set of
tiles~$T$ with the following properties:

{\leftskip 20pt\parindent 0pt
$(i)$ for each tile~$\tau$ of~$T$, exactly two edges of~$\tau$ are marked;
the mid-points of these edges define an arc in~$\tau$ which we call
a {\bf path element};

$(ii)$ there is a solution of the tiling problem of~$T$ such that
the path elements are abutted into a single path.
\par}

   Note that due to the condition~$(i)$, the path is not a cycle.
Also note that, in the formulation of the problem, the set~$T$ does not 
define an initial tile. 

   In this case, the path defined by the tiling is called a {\bf uniform
plane-filling path}. Note that, both for regular grids of the Euclidean or the 
hyperbolic planes, it is not difficult to construct paths of the plane 
which visit each tile exactly once, when starting from a distinguished tile
of~$T$.

   The {\bf strong} plane-filling property consists in finding
a finite set satisfying the simple plane-filling property together
with an additional condition:

{\leftskip 20pt\parindent 0pt
$(iii)$ for any solution of the tiling problem of~$T$, the path elements
are abutted into a single path.
\par}

   In other words, in case of the strong property, any solution for tiling
the plane with~$T$ defines a uniform plane-filling path. Note that the 
plane-filling path defined in this way may then be different from one 
solution to another.

   Our construction defines a uniform plane-filling path in the
hyperbolic plane and the generating finite set of tiles almost possesses 
the strong plane-filling property. All its solutions generate a plane-filling
path, expected in one case. In that case, the path elements 
constitute a countable family of disjoint infinite paths whose union visits 
each tile exactly once. However, this solution can be seen as a limit case 
of the other solutions. All these paths, but two ones, $\pi_1$~and~$\pi_2$
can be joined at infinity and the new path and~$\pi_1$ join~$\pi_2$ at 
infinity. In some sense, these paths are the trace of a unique path also 
visiting the points at infinity. 

   As mentioned in our abstract, the Euclidean plane satisfies the strong
plane-filling property, established by J. Kari, see~\cite{jkari94}. 
Accordingly, the paper shows that the hyperbolic plane also possess the
simple version of the property. Call {\bf ternary heptagrid}, 
see~\cite{mmbook1}, the tiling of the hyperbolic plane
based on the regular heptagon with the angle~$\displaystyle{{2\pi}\over3}$.
We prove the following property.

\begin{thm}\label{plane_fillka}
There is a uniform plane-filling path for the tiling of the ternary
heptagrid of the hyperbolic plane.
\end{thm}

   We also prove another property:

\begin{thm}\label{CA_fillka}
There is a cellular automaton on the ternary heptagrid
which constructs a uniform plane-filling path in infinite time.
\end{thm}
\vskip 5pt
   The existence of such a path was required for proving a property on 
cellular automata. Say that a cellular automaton is reversible if and only 
if its global transition function is bijective and also defined by a 
cellular automaton. From the plane-filling property which he established, 
J. Kari proved that it is undecidable to decide whether a cellular automaton 
on the Euclidean plane is reversible or not, see~\cite{jkari94}. The similar
question for cellular automata in the hyperbolic plane is open.
  
   Our construction relies on the construction which we defined in
\cite{mmarXivb,mmarXive} in order to establish that the domino problem
is undecidable in the hyperbolic plane. We very sketchilly remind
this construction in section~2, mainly reminding what is needed for
the present construction.

   In section~3, we indicate the construction of new triangles, the
{\bf mauve} triangles which we shall use for guiding the travel of the
path. In section~4, we describe the construction of a uniform plane-filling
path. We prove that the hyperbolic plane almost possesses the strong 
property. We also prove that there is a cellular automaton on the 
tiling~$\{7,3\}$ which is able to construct a uniform plane-filling path, 
of course, in infinite time.

   In section~5, we give an application to an algorithmic construction 
of a Peano curve in the hyperbolic plane. However, we have a simpler
construction of such a curve which we shall give in a forthcoming paper.

\section{The underlying construction}

   The tiling which we use is based on the {\bf ternary heptagrid},
the tessellation~$\{7,3\}$ of the hyperbolic plane. We remind that it is
generated by reflection of a regular heptagon with interior 
angle~$\displaystyle{{2\pi}\over3}$ in its edges and, recursively, of
the images in their edges. An illustration of this tiling is given by
figure~\ref{splittil_73}, and we refer the reader to~\cite{mmbook1}
for the properties of the tiling used in this paper.

\vskip 10pt
\setbox110=\hbox{\epsfig{file=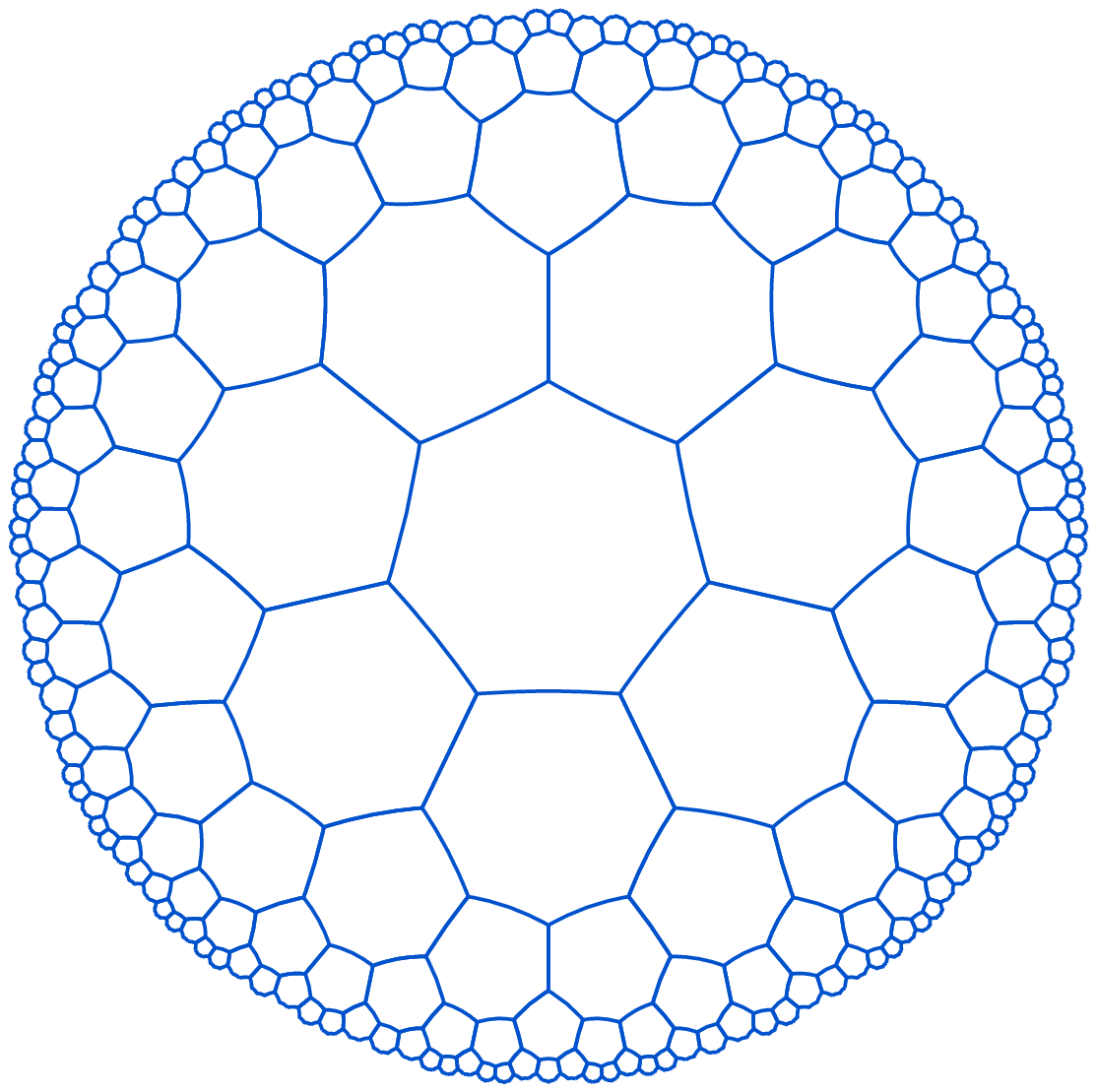,width=210pt}}
\ligne{\hfill
\PlacerEn {-290pt} {-90pt} \box110
}
\begin{fig}\label{splittil_73}
\leurre
The tiling $\{7,3\}$ of the hyperbolic plane in the Poincar\'e's disc
model.
\end{fig}
\vskip 10pt
   In this tiling, we introduced an auxiliary tiling, the {\bf mantilla},
which was first defined in \cite{mmarXiva} and which is used in 
\cite{mmarXivb,mmarXive} to prove the undecidability of the tiling problem
in the hyperbolic plane. 

   The construction of the mantilla is based in fixing rules to assemble
two kinds of tiles: the {\bf centres} and the {\bf petals}, the tiles~$\alpha$
and~$\beta$ of figure~\ref{momo_petals}, respectively. By numbering the 
edges of the centres from~1 up to~7, we prevent the centres to tile the 
plane by themselves, alone. It is needed to put petals around them.

\setbox110=\hbox{\epsfig{file=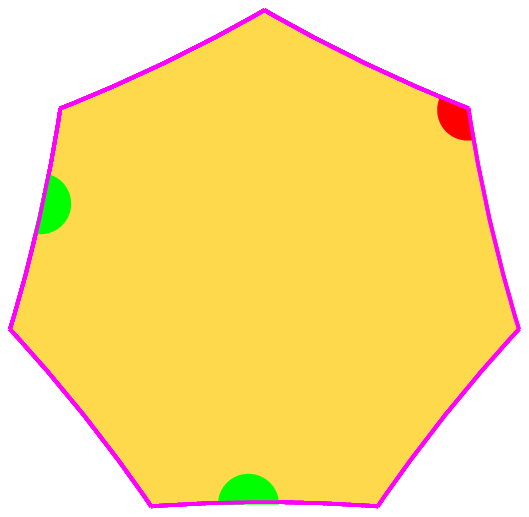,width=200pt}}
\setbox112=\hbox{\epsfig{file=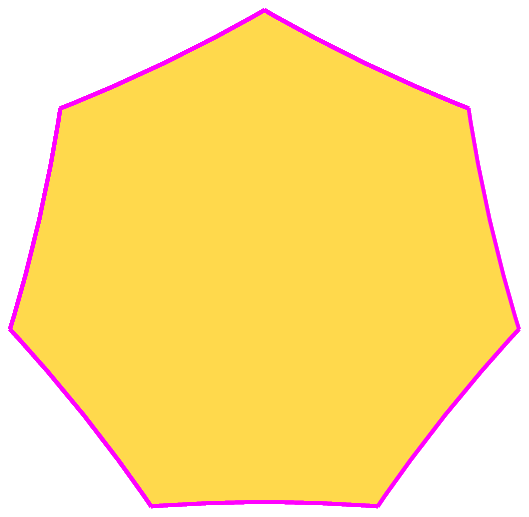,width=200pt}}
\ligne{\hfill
\PlacerEn {-350pt} {-80pt} \box112
\PlacerEn {-210pt} {-70pt} {$\alpha$}
\PlacerEn {-215pt} {-80pt} \box110
\PlacerEn {-75pt} {-70pt} {$\beta$}
}
\begin{fig}\label{momo_petals}
\leurre
Left-hand side: the tile for the centres of the flowers. Right-hand side:
the tile for the petals.
\end{fig}
\vskip 10pt
   Now, we can rule the way in which petals are put around a centre,
making a figure which we call a {\bf flower}. We define four types
of flowers. Figure~\ref{til_mantilla} indicates three types of them:
the types~$F$, $G$ and~{\bf 8}. Now, the type~$G$ has two variants, which
we call~$G_\ell$ and~$G_r$, respectively, which are in some sense symmetric.
The distinction is a consequence of the numbering of the edges which is
the same in both cases.
  
\vskip 10pt
\setbox110=\hbox{\epsfig{file=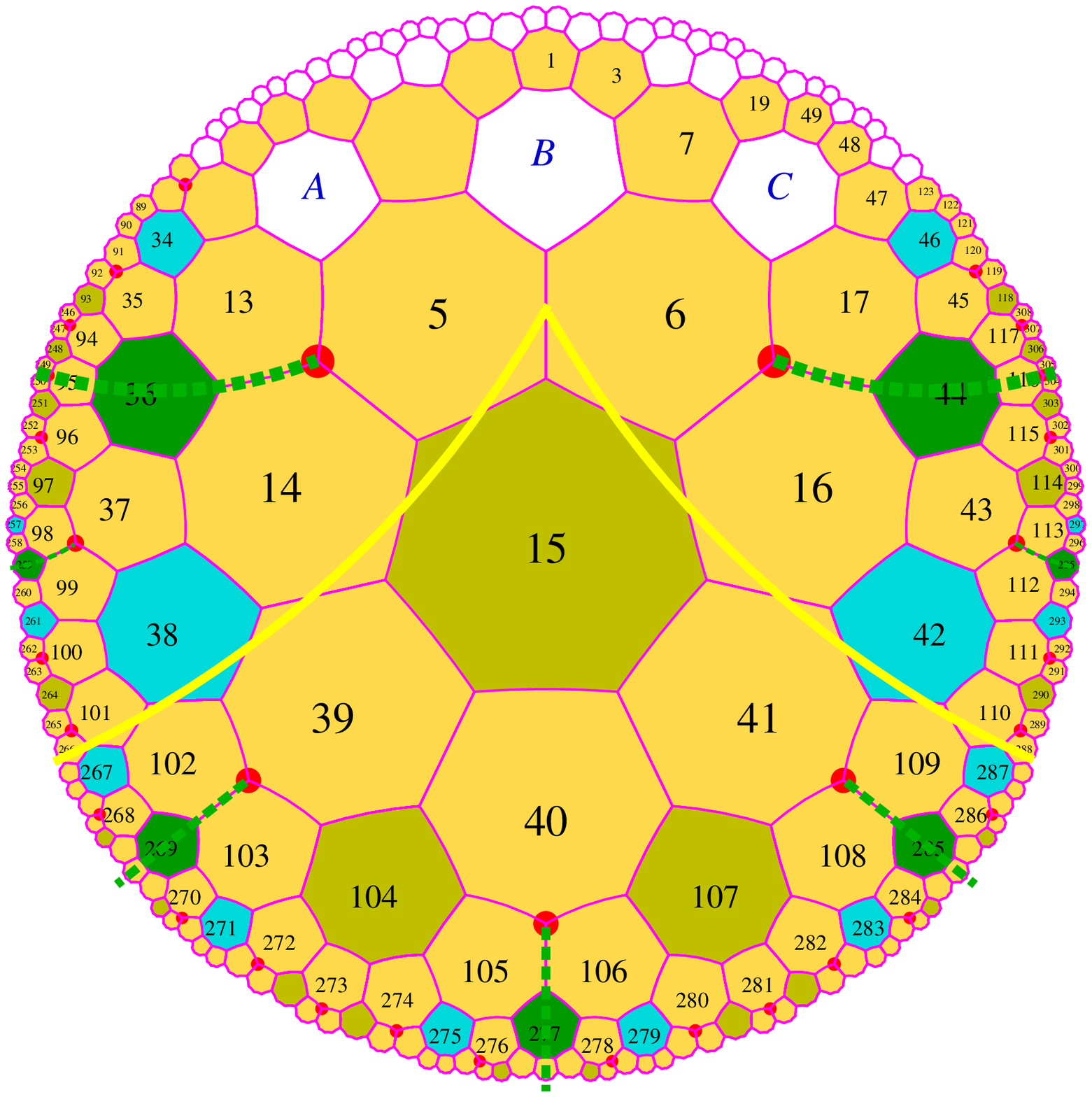,width=100pt}}
\setbox112=\hbox{\epsfig{file=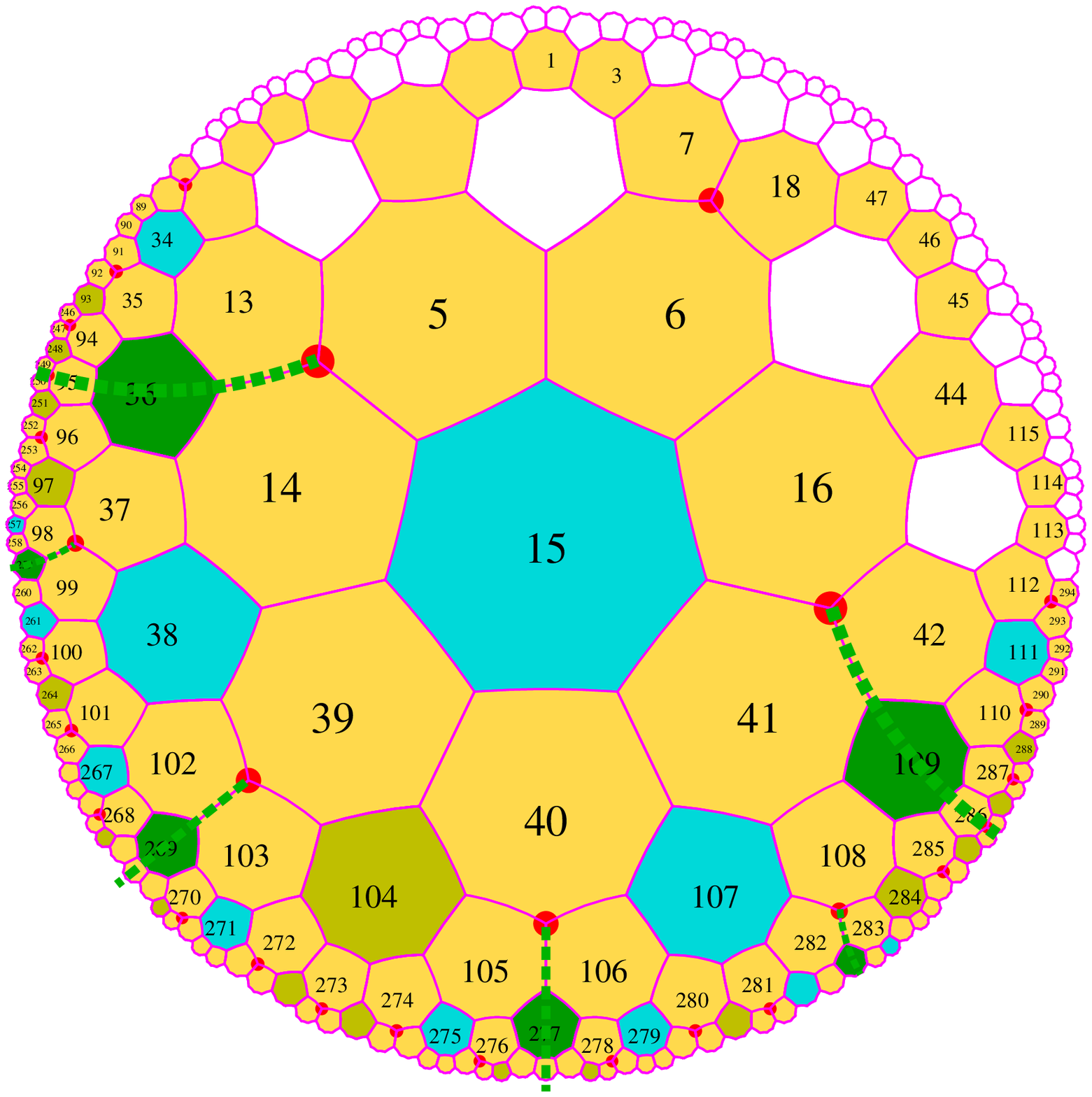,width=100pt}}
\setbox114=\hbox{\epsfig{file=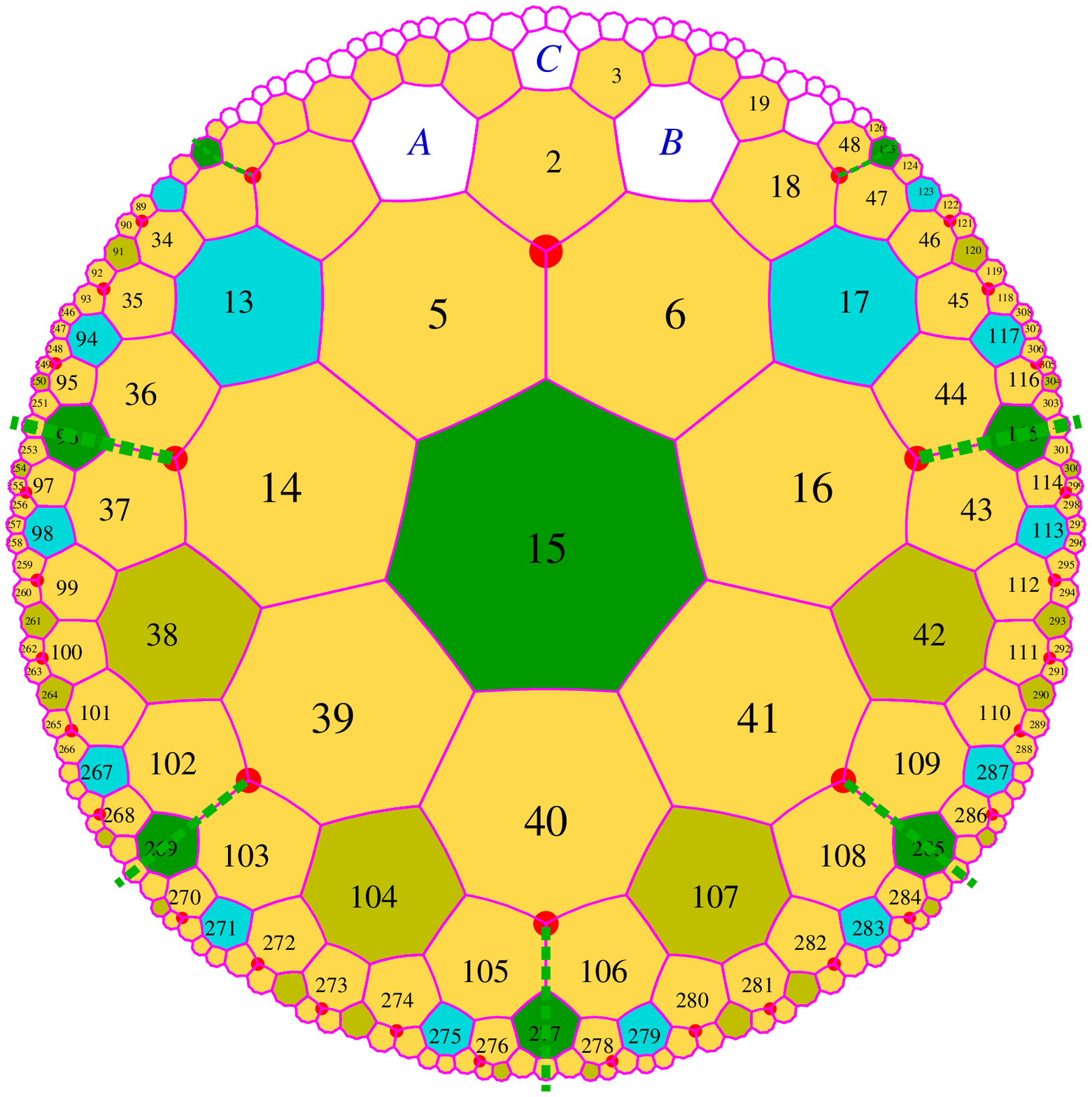,width=100pt}}
\ligne{\hfill
\PlacerEn {-345pt} {-60pt} \box110
\PlacerEn {-255pt} {-55pt} {$F$}
\PlacerEn {-225pt} {-60pt} \box112
\PlacerEn {-135pt} {-55pt} {$G$}
\PlacerEn {-105pt} {-60pt} \box114
\PlacerEn {-15pt} {-55pt} {\bf 8}
}
\begin{fig}\label{til_mantilla}
\leurre
Splitting of the sectors defined by the flowers. From left to right:
an $F$-sector, $G$-sector and {\bf 8}-sector.
\end{fig}

   These figures also display the way in which each sector determined by
a flower is split in such a way that in each sector, the complement of its
defining flower can be expressed in $F$- and~$G$-sectors, with the help of
half {\bf 8}-sectors. The green rays of the three pictures in
figure~\ref{til_mantilla} indicate this splitting. This defines a
recursive process to generate the mantilla. The algorithm is deterministic
when we proceed downwards and it is non-deterministic when we proceed
upwards.
 
   A last ingredient consists in introducing {\bf isoclines}, which play the 
r\^ole of horizontals in the Euclidean plane. The levels are illustrated by
figure~\ref{mant_levels}, below. They are defined by fixing the {\bf 8}-centres
as black nodes in the sense given to the nodes of Fibonacci trees, 
see~\cite{mmbook1,mmarXivb}. Four other cases appear outside those
indicated in figure~\ref{mant_levels}, we refer the reader to~\cite{mmarXivb}.

\vskip 10pt
\setbox110=\hbox{\epsfig{file=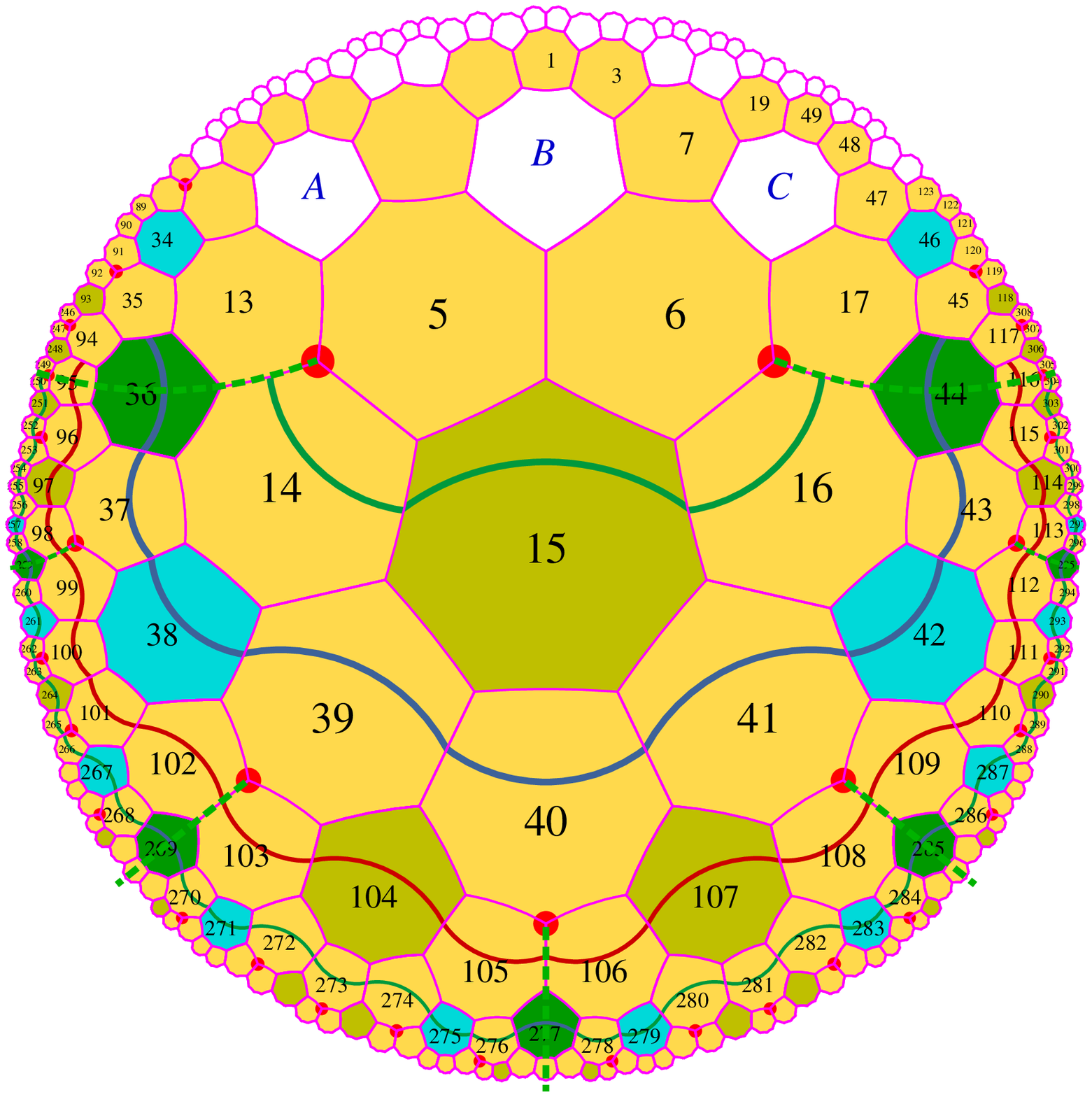,width=100pt}}
\setbox112=\hbox{\epsfig{file=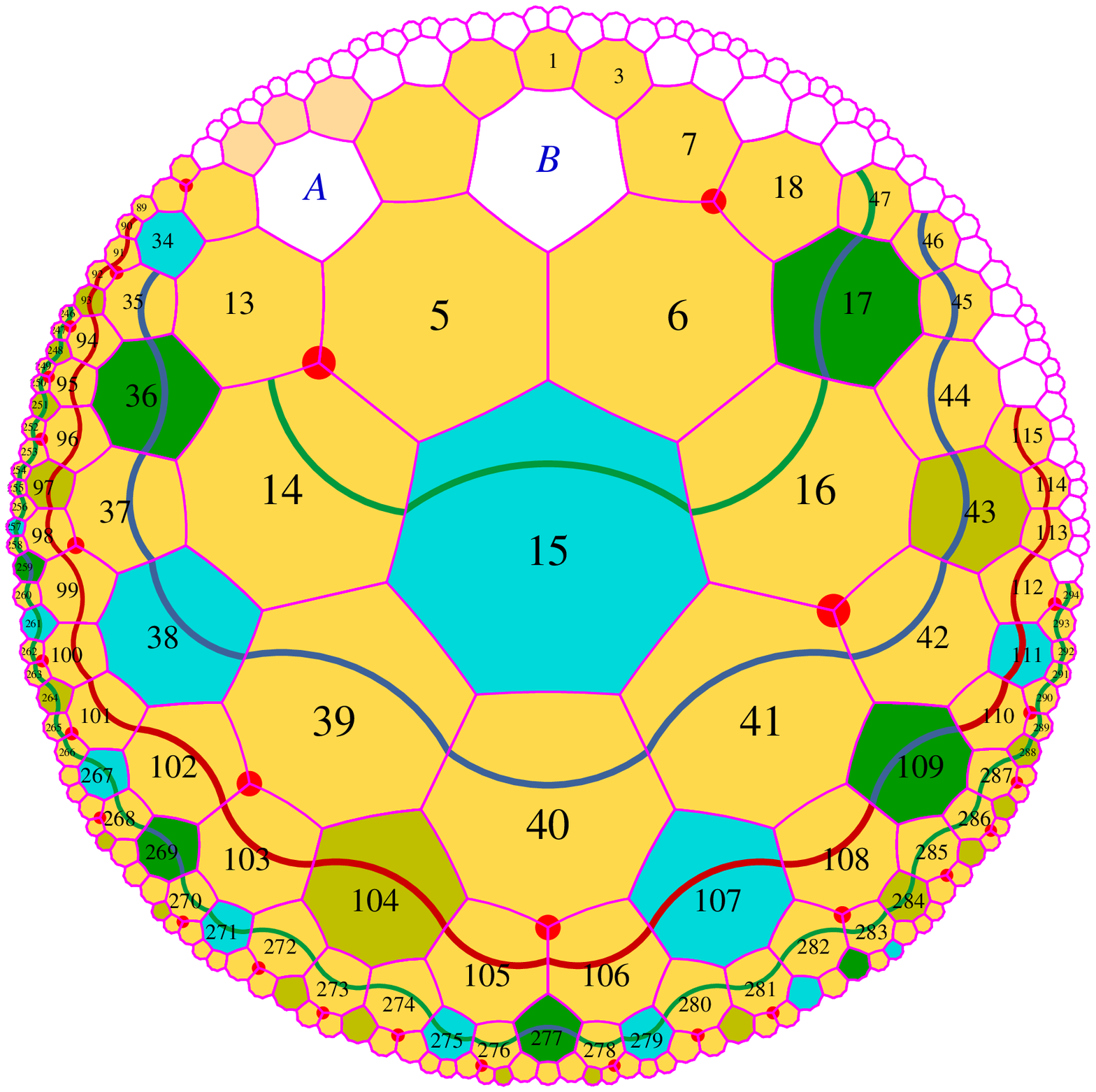,width=100pt}}
\setbox114=\hbox{\epsfig{file=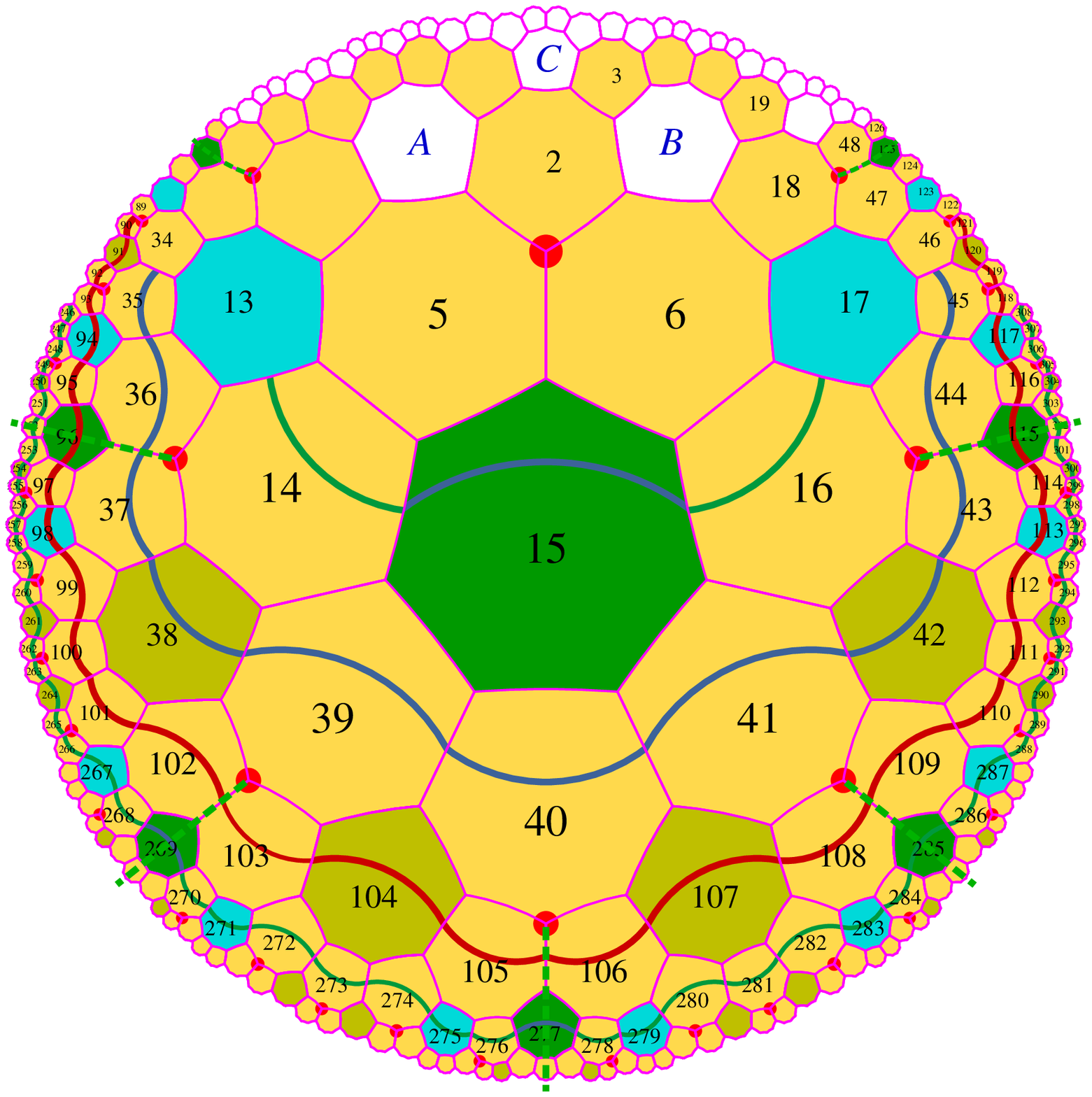,width=100pt}}
\ligne{\hfill
\PlacerEn {-345pt} {-60pt} \box110
\PlacerEn {-255pt} {-55pt} {$F$}
\PlacerEn {-225pt} {-60pt} \box112
\PlacerEn {-135pt} {-55pt} {$G$}
\PlacerEn {-105pt} {-60pt} \box114
\PlacerEn {-15pt} {-55pt} {\bf 8}
}
\begin{fig}\label{mant_levels}
\leurre
Splitting of the sectors defined by the flowers. From left to right:
an $F$-sector, $G$-sector and {\bf 8}-sector.
\end{fig}

   These isoclines are very important: they are the basis of the construction
of the {\bf interwoven triangles} which we need for proving 
theorem~\ref{plane_fillka}. These triangles were introduced in
\cite{mmarXivb,mmarXive} in order to prove the undecidability of the
tiling problem in the hyperbolic plane.

   Now, we sketchilly indicate how to construct these triangles in the 
Euclidean plane. First,
we have a line of light blue equal triangles, as it can be seen in 
figure~\ref{interwoven_fig}. They are isoceles and their main heights are 
supported by the same line, the {\bf axis}. Note that triangles with thick 
edges alternate with triangles with thin edges. 
Following~\cite{mmarXivb,mmarXive}, we call {\bf phantoms} the triangles
with thin edges. The triangles and phantoms which we just described
constitute those of the generation~0. Now, the colours of the generation 
will alternate between red and blue, 
\ligne{\hfill}
\setbox110=\hbox{\epsfig{file=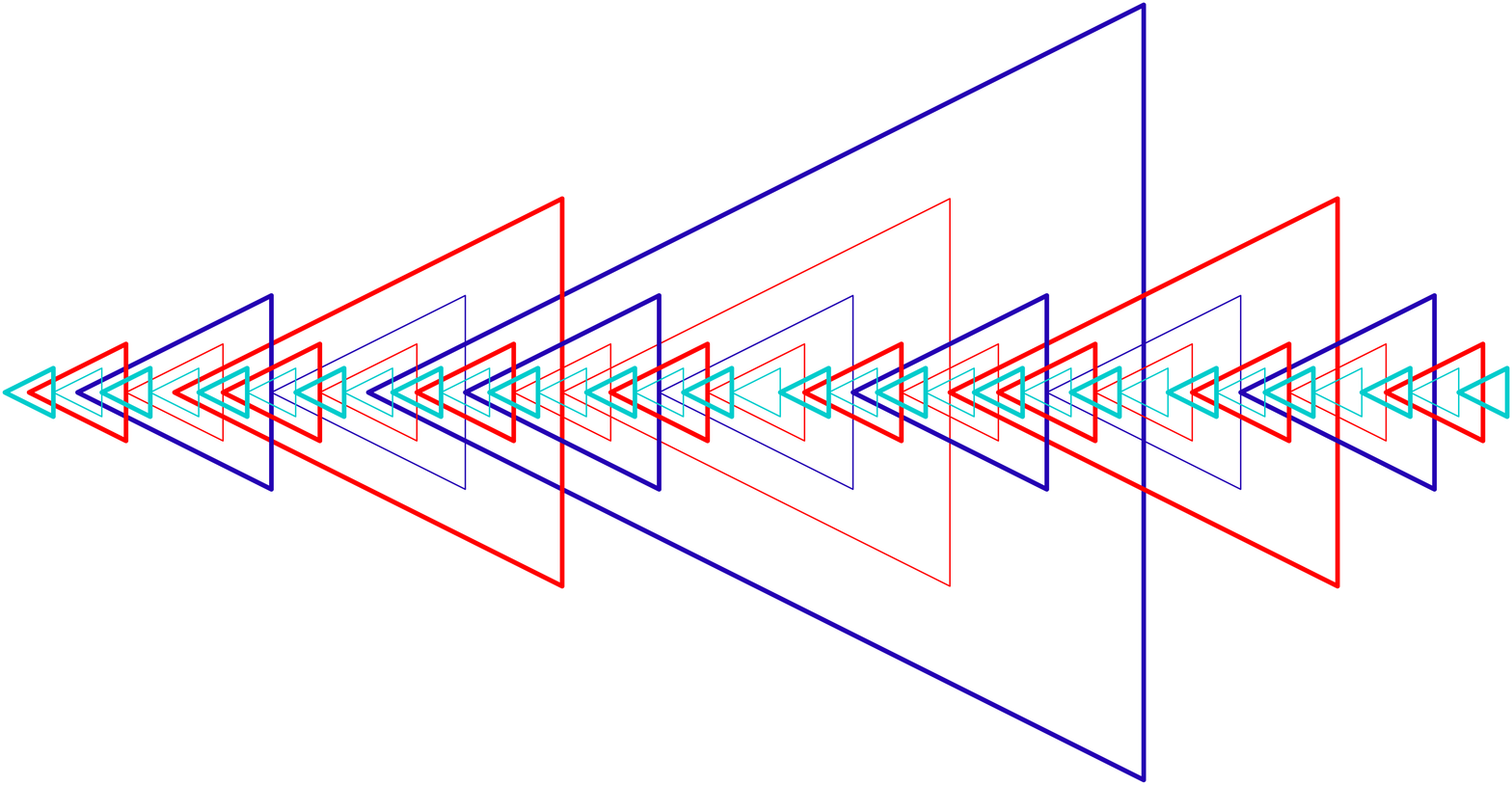,width=300pt}}
\ligne{\hfill
\PlacerEn {-330pt} {0pt} \box110
}
\vspace{-10pt}
\begin{fig}\label{interwoven_fig}
\leurre
An illustration for the interwoven triangles.
\end{fig}
which will be a medium blue, and say 
that red and blue are {\bf opposite} to each other. Assume that we 
constructed the 
generation~$n$. We fix a triangle of the generation~$n$ and, at the 
mid-point of its main height, on the axis, we put the vertex of a 
triangle of the opposite colour, with respect to that of the generation~$n$.
Then, we construct an isoceles triangle~$T$ whose height is supported by 
the axis, in such a way that its basis crosses the main height of the next 
triangle of generation~$n$. For simplicity, we may assume that the 
legs of~$T$ are parallel to the corresponding legs of the triangles of 
generation~$n$ which are all equal and whose legs are also parallel. We 
replicate~$T$ by shifts along the axis in such a way that we obtain an
alternation of triangles and phantoms of the same colour as~$T$ and such
that a vertex of a phantom is on the mid-point of a basis of a triangle 
and a vertex of a triangle is on the mid-point of a basis of a phantom. 
These triangles and phantoms, $T$ being included, constitute the 
generation~$n$+1. Figure~\ref{interwoven_fig} illustrates
this point. 
\vskip 5pt
   Our last step is to implement these triangles in the hyperbolic plane.
We are faced with three problems.

   First, the choice of the place of the triangles with respect to
those of the previous generation generates a continuous number of solutions.
Let us call {\bf infinite model} a given way to fix the successive generations.

   The second is to define what will be the supports of the triangles 
and the phantoms and what will be their axis. We have to leave 
the precise details to the quoted papers, \cite{mmarXivb,mmarXive}, but here,
we still give an idea of the situation. 

   First we define the supports of the triangles and the phantoms. The 
isoclines which we defined are periodically numbered from~$0$ up to~19, the 
incresing numbers going downwards. The number~20 appears for technical 
reasons which are clearly explained in~\cite{mmarXivb}. Now, the candidates 
for the support of the triangles are defined by Fibonacci trees, 
see~\cite{mmbook1}, rooted at the $F$-son of a $G$-flower on an 
isocline~0, 5, 10 or~15. Now, not all the indicated such nodes, call 
them {\bf seeds}, are allowed to generate a tree, in which case we say that 
the seed is {\bf active}. However, all seeds on an isocline~0 are active. 
But, for the others, they are active only if they are inside a tree rooted 
at an active seed. This induces a tree of the active seeds on the 
isoclines~5, 10 and 15 which have the same ancestor inside a given tree 
rooted at an isocline~0. Such branches and their upward continuations are 
called {\bf threads}. The legs of triangles are supported by the extremal 
branches of the trees rooted at active seeds and the r\^ole
of the axis is played by the threads. This induces many problems.

   It may happen that a thread traverse the hyperbolic plane. If this happens,
the corresponding threads do coincide, starting from a certain point. We call
such threads {\bf ultra-threads}. All the other threads have the structure 
of a ray. Now, the existence of ultra-threads or not depends on the 
particular mantilla which we constructed. 

   The other point is that we better control the situation if the triangles
and the pantoms of the same generation but on different threads have their 
vertices and bases on the same isoclines. In this case, we say that the 
triangles are {\bf synchronized}. Synchronizing the triangles, of course, 
also the phantoms, boils down to consider that each thread implements the 
same model of interwoven triangles. Now, something must be made more clear. 
We can realize a whole infinite model along an ultra-thread. But, as a 
thread is bounded from above, this is not possible for an ordinary thread. 
In fact, we have to study what happens in an infinite model if 
we introduce a cut: we fix a line~$\lambda$, orthogonal to the axis, we 
erase all triangles whose vertex is on the left-hand side of~$\lambda$ and
we keep all of them which are on its right-hand side.

   In \cite{mmarXivb,mmarXive}, we proved that by observing these constraints,
we can obtain the synchronization of all the implementations of the cuts
defined by the threads of the same infinite model. Note that as we have
continuously many different realizations of the mantilla and continously
many different realization of an infinite model, we have in fact continously
many implementations of the interwoven triangles.

   At this point, let us note that if we could fix the infinite model followed
by the interwoven triangle, we could skip the next section and directly go
to the following one, making it much more simple. But one model, which 
cannot be avoided, requires the solution which we define in the mauve 
triangles.

   Before turning to what is introduced for proving theorem~\ref{plane_fillka},
we insist on a particularity of the implementation: a triangle always
contains several triangles of the previous generation on the same set of 
isoclines, which is another aspect of the synchronization.

\section{The mauve triangles}

   Now, we turn to the construction of the regions which control the
path which is defined in the next section.

\subsection{Construction of the mauve triangles}

   To this purpose, we keep the red triangles only, but we keep in mind the
red phantoms generated by their bases, as they play a r\^ole in the 
construction. 

   Now, to the red triangles, we supperpose new triangles, which we call
the {\bf mauve} triangles. Each vertex of a red triangle is also the vertex
of a mauve triangle and conversely. The legs of the mauve triangle are
supported by the legs of the red triangles, but they go further, on the same
extremal branch of the tree which defines the red triangle. The legs are
stopped by the next isocline supporting a vertex of a red triangle of the
same generation. In some sense, the length of the height of a mauve triangle
is twice the length of the height of the red triangle sharing its vertex.

   It is not difficult to construct the mauve triangles from the 
red triangles. Consider a red triangle~$T$. The mauve triangle~$M$ defined 
by~$T$ is constructed as follows. The vertex of~$M$ is that of~$T$ and its
legs follow those of~$T$. When a leg of~$M$ arrives to the corner of~$T$,
this corner sends a {\bf purple} signal along the basis of~$T$ in the 
direction of the other corner: this can easily be determined by the tiles
which materialize the corners of~$T$. When this signal reaches the
first vertex of a phantom~$P$, necessarily a red one and of the same generation
of~$T$, it goes on the leg of the phantom which is
on the same side as the corner of~$T$ which it has left and it goes down
along the leg of~$P$ until the corner of~$P$. Then, it goes on its way on
the same isocline as the basis of~$P$ but on the direction which goes 
outside~$P$. The purple signal goes on until it meets the mauve leg which
has continuated its way on the extremal branch of the tree supporting~$T$,
from the corner from which the purple signal originated.

   From the  point of view of the tiling, it is important to notice that 
the purple signal cannot be generated by a phantom which would be 
internal, in this sense that it would have a phantom on each side whose 
vertices lie on the basis of the same triangle. To realize this, it is
enough to give a {\bf laterality} to the purple signals: a purple signal
inherits the laterality of the triangle corner from which it is originated.
It is enough to forbid a joining tile to prevent the emission of a
purple signal by a wrong phantom. Now, as a horizontal purple signal 
running on the isocline of the phantom and outside it must meet the leg of 
a mauve triangle, on the inside part of the leg, always clear from the 
tiles, see \cite{mmarXivb,mmarXive}, if the emitting triangle does not 
exist, which may happen, even if the phantom exists, then the purple signal 
will meet a leg of phantom of the opposite laterality: it is easy to rule 
out this.

\vskip -60pt
\setbox110=\hbox{\epsfig{file=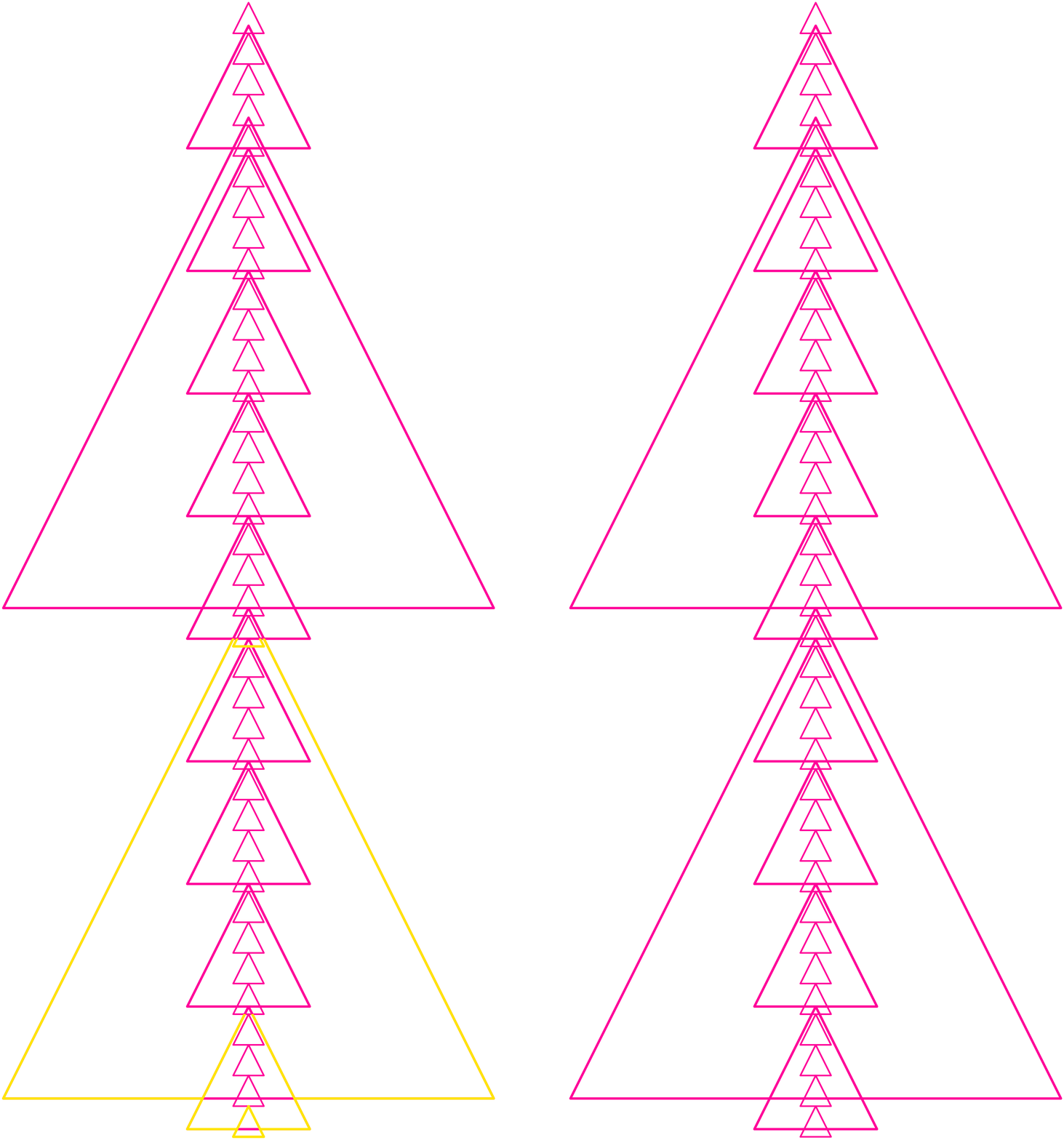,width=300pt}}
\ligne{\hfill
\PlacerEn {-330pt} {0pt} \box110
}
\vspace{-10pt}
\begin{fig}\label{lesmauves}
\leurre
An illustration for the mauve triangles.
\end{fig}

   Now, the purple signal has only a construction purpose. As it plays no 
more r\^ole, we shall forget it in the representations of the mauve triangles,
see figure~\ref{lesmauves}. 
 
   Presently, let us indicate the properties of the mauve triangle.

\subsection{Properties of the mauve triangles}

   Using the terminology of the interwoven triangles, 
see \cite{mmarXivb,mmarXive}, we say that the set of isoclines crossed by
a mauve triangle, the basis and the vertex being included, defines
the {\bf latitude} of the mauve triangle. Also, we know that red triangles
have an odd index in the generations of the interwoven triangles. We shall
say that a mauve triangle associated to a red triangles of generation~$2n$+1
is of generation~$n$. The first property is very
important for the following:

\begin{lem}\label{cover}
Let~$\tau$ be a tile of the tiling. Then for any non-negative~$n$, 
there is a mauve latitude $\Lambda$ of this generation such
that $\tau\in\Lambda$. And then: either $\tau$ falls within a mauve 
triangle of generation~$n$ in this latitude or $\tau$ falls outside
two consecutive mauve triangles of generation~$n$ and of the 
latitude~$\Lambda$ and in between them.
\end{lem}

   This property follows immediately from the fact that the latitude of
a mauve triangle exactly covers that of the corresponding red triangle
and the following latitude of red phantoms.

   However, there is a price to pay to this: the red triangles are
either disjoint or embedded. Mauve triangles do intersect from one 
generation to another. Fortunately, this intersection is not that big
and we characterize it in the following statement.

\begin{lem}\label{overlap}
A mauve triangle~$T$ of positive generation~$n$ intersects mauve triangles
of generation~$n$$-$$1$, and it possibly intersects one mauve triangle
of generation~$n$$+$$h$$+$$1$, with $h\geq 0$. When the intersection occurs, i
the legs of~$T$ cut the basis of the mauve triangle of the higher 
generation at a point which is on the mid-distance line of the phantoms of 
generation~$2(n$$+$$h$$+$$1)$$+$$1$ which share 
their basis with that of~$T$. Call {\bf low point} this point on the legs 
of~$T$. The basis of~$T$ is cut by the legs of mauve triangles of 
generation~$n$$-$$1$ at their low points.
\end{lem}

   The proof is easy and it comes from the relations of red triangles of
consecutive generations. Representing the first three generations,
figure~\ref{lesmauves} illustrates this property.

\subsection{Determination of the low points}

   As we shall see in the next sections, the low points of a mauve triangle
play an important r\^ole. Let us show that they can be determined from the
tiles themselves.

   Consider a mauve triangle~$T$. Let $R$~be the red triangle which shares
its vertex with that of~$T$ and let $P$~denote both, the leftmost and
rightmost phantoms generated by the basis of~$T$. From the above definition, 
we know that the low points of~$T$ lie on the isocline which supports the 
mid-distance line of~$P$. Now, the leg of~$P$ which is on the same side
as the closest mauve leg of~$T$ are covered by the purple signal with the 
laterality of the signal coinciding with that of the leg. On the part of 
this signal which runs on the leg of~$P$, the mid-point is easily found. 
Accordingly, the purple signal sends a {\bf dark purple signal} on the 
corresponding isocline, outside~$P$. In between the leg of~$P$ and the 
leg of~$T$, the dark purple signal will meet mauve legs of triangles of 
smaller generations, see figure~\ref{lesmauves}. To avoid problems 
connected with possible nestings of such triangles, the purple dark 
signal looks at the laterality of the first mauve leg it meets: if it is 
of the same laterality as his own one, he found the appropriate leg. If 
not, it climbs along the mauve triangle until it meets its vertex and 
goes down on the other side: by induction, it is assumed that the low 
points of mauve triangles of previous generations have been determined. 
Assume that this is the case. Then the dark purple signal goes on along 
the right isocline, avoiding smaller mauve triangles possibly contained on 
those he jumped over. Now, it will meet a first mauve leg of its laterality 
which will be the expected one and so, it will determine the low point of 
this leg. And so, as the low point for the mauve generation~0 is easy to 
determine because it contains no mauve triangle, this process works and
it can easily be implemented with finitely many tiles. Note that this 
process is similar to the one which we used in~\cite{mmarXivb,mmarXive}, 
in order to synchronize horizontal signals travelling on certain isoclines.

   Now, we can turn to the construction of the path.

\section{A uniform plane-filling path}

   Until the last sub-section, we assume that all the mauve triangles 
of the tiling are {\bf finite}. In the last section, we shall see what
happens when this is no more the case.

   The construction of the path is based upon two basics patterns which
we now define.

\subsection{The guidelines}

   The idea is to look at things globally, at the level of latitudes
of mauve triangles. We have to check that we can construct the path as the 
result of an algorithmic process, infinite in time, punctuated by times~$t_k$
in such a way that at time~$t_{k+1}$ we fill up more space in the latitudes
already visited up to time~$t_k$ and that at time~$t_{k+1}$, we access
to latitudes of higher generation with respect to those accessed up to
time~$t_k$.

   We shall look at the tiling by making a rather rough approximation which
will turn out to be good enough. This consists in looking at the mauve 
triangles as if there were no overlapping between different generations: we 
can look at latitudes as over disjoint or embedded. We shall call this the 
{\bf first approximation}.

   At the level of the latitudes, in first approximation, we have two
figures: the triangle and the trapeze, see figure~\ref{tri_trap}. 
The triangle is a mauve triangle
whose height is that of the latitude and the trapeze is the part of a latitude
which lies in between two consecutive mauve triangles of the latitude and of
its generation, also delimited by the isoclines going through the vertices
and the bases of these triangles.

   There are two {\bf access} of the path into the triangle or into
the trapeze. They will be called {\bf entries} or {\bf exits}, depending
on the way we look at the path which, by construction, is not oriented. 
In fact, looking at the figures from the left to the right, we can
define two displays for the access. There is the {\bf ascending} one:
an access on the lower left-hand side corner and the other access at the
vertex, for the triangle, looking to the right, and on the upper right-hand
side corner for the trapeze. There is a {\bf descending} display:
an access on the lower right-hand side corner and the other access at the
vertex, for the triangle, looking to the left, and on the upper left-hand
side corner for the trapeze.

   Note that the descending figures of one kind match with the ascending ones
of the other kind, also provided that they exactly fit within the same 
latitudes.

   Now, the path cannot remain for ever within a given latitude of the 
mauve triangles. This is a consequence of lemma~\ref{cover}. Now, the 
trapezes and triangles which we define are bigger and bigger. If we fix a 
tile~$\tau_0$ once for all, then after a time $t_{k_0}$, the path 
completely contains a ball of radius~$n$ around~$\tau_0$ at time~$t_{k+n}$. 
It is enough to define the $t_k$'s by this condition which is satisfied 
by lemma~\ref{cover}: note that a trapeze is much much bigger than a 
triangle of the same latitude.

\setbox110=\hbox{\epsfig{file=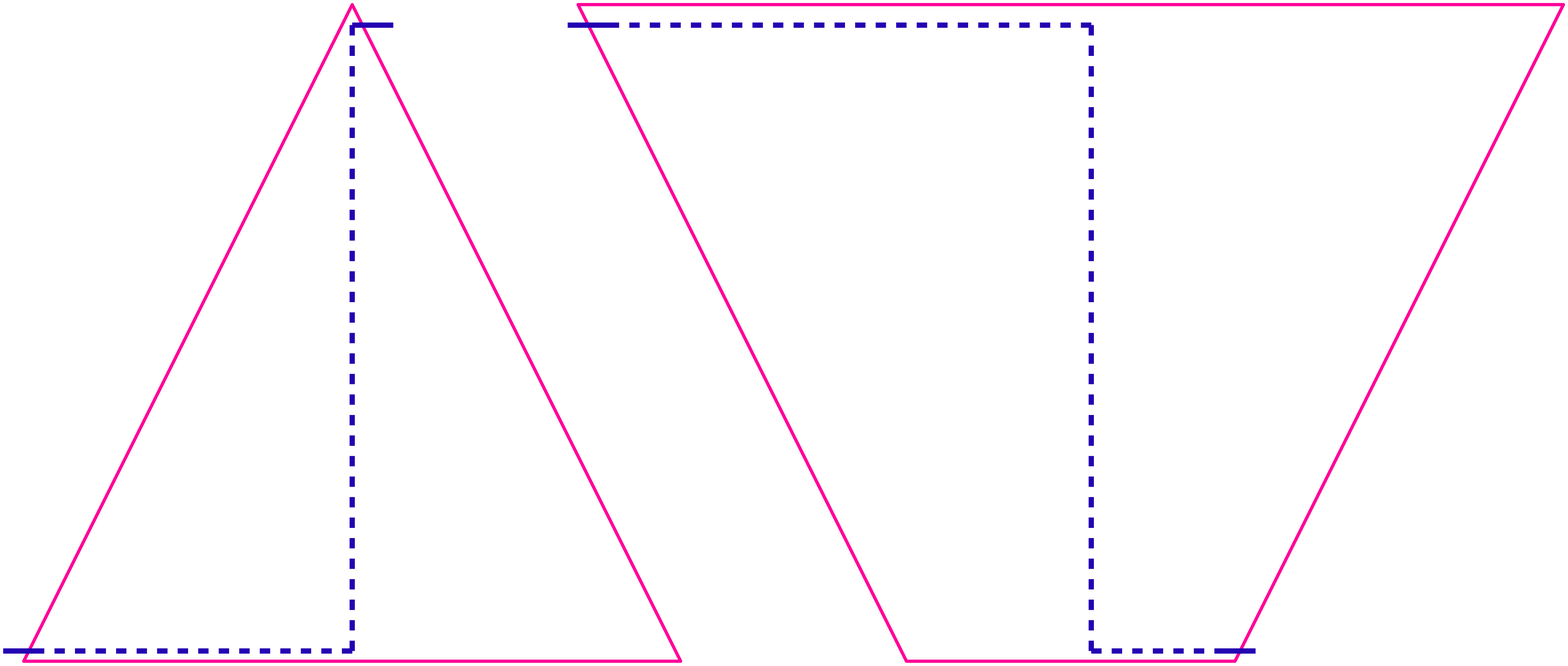,width=300pt}}
\vskip 10pt
\ligne{\hfill
\PlacerEn {-330pt} {-10pt} \box110
}

\begin{fig}\label{tri_trap}
\leurre
The basic figures: triangle and trapeze within a latitude.
\end{fig}

   Now, we have to define the internal structure of the triangles and trapezes
above defined. For this, we notice that inside a triangle, there
are four latitudes of mauve triangles of the previous generation, and, in
first approximation, these latitudes are disjoint.

   We shall look at the way we can fulfill the requirement posed upon the 
triangles and the trapezes. For this, we shall introduce an intermediate
picture which we call the {\bf quadrangle}, as the word {\it rectangle}
would be misleading in this context. The advantage of this figure is
that we can look at it as the trace of a latitude either inside a triangle
or inside a trapeze, see figure~\ref{quadrangle}. We have two kinds
of such quadrangles, corresponding to the main motion of the path in filling
up this region. One version is the descending one, see picture~$(d)$ in
figure~\ref{quadrangle}. The second version is the ascending one, see 
picture~$(a)$ in the same figure. In the decomposition of a quadrangle, 
we again find triangles and trapezes, but of the previous generation, which 
are smaller. The decomposition is repeated until the generation~0 is reached.

   Three precisions must be given about the quadrangles, see 
figure~\ref{quadrangle}. 

   First, a quadrangle occurs both in a trapeze
and a triangle. Note that the lateral sides of a quadrangle determine
where it is in a given latitude. If the lateralitites of these sides
are identical, we are inside a triangle and outside a smaller one. If the 
lateralities are different, we have to look at whether they correspond 
to the position of these borders with respect to the region which they 
delimit. If the lateralities of the borders define their position, we 
are inside a triangle of the considered latitude. If the lateralities of 
the borders are opposite to their actual position, we are in a trapeze 
of the considered latitude.

   The second point is about the dotted lines in figures~\ref{tri_trap},
\ref{quadrangle} and~\ref{quadrangle_deux}. In these figures, the blue 
dotted lines represent zig-zags going from one vertical border to the other 
while following an isocline. The zig-zag runs over all the isoclines of the 
considered areas, one after the other. In figure~\ref{quadrangle}, the 
green dotted line represents a path along an isocline. The blue dotted-line 
is a zig-zag which is bordered by the green dotted line and by the concerned 
borders: they are legs of triangles, but this time, the place of the
region bordered by the legs is defined by their lateralitites.
\setbox110=\hbox{\epsfig{file=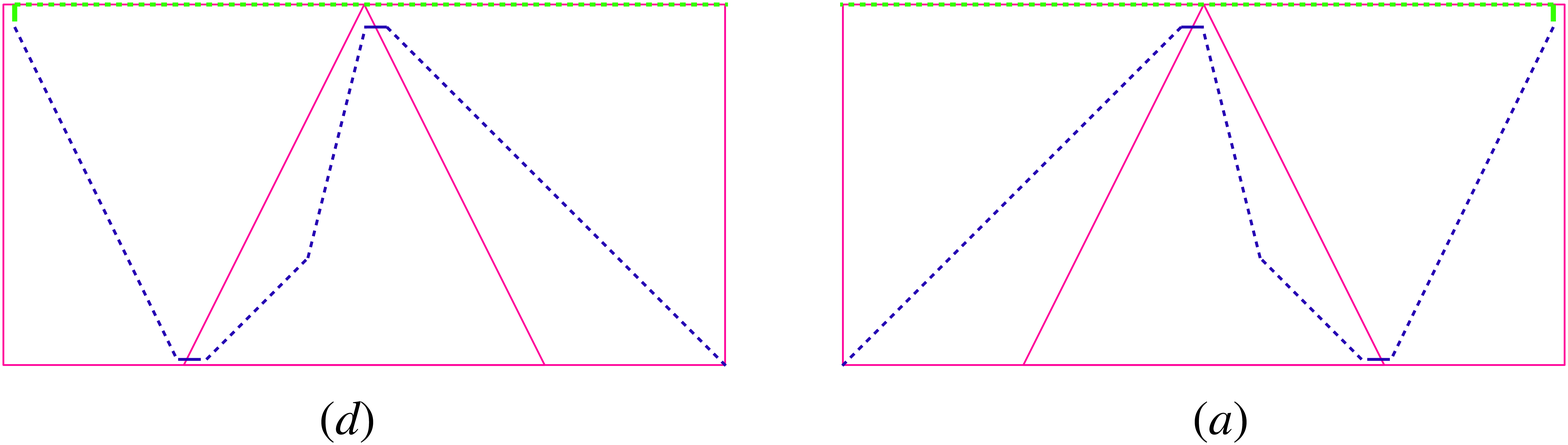,width=350pt}}
\ligne{\hfill}
\vskip -35pt
\ligne{\hfill
\PlacerEn {-350pt} {-40pt} \box110
}
\vspace{-30pt}
\begin{fig}\label{quadrangle}
\leurre
The splitting of the slices.
\vskip 0pt\parindent 0pt
On the left-hand side: the descending case. On the right-hand side:
the ascending case.
\end{fig}

\setbox110=\hbox{\epsfig{file=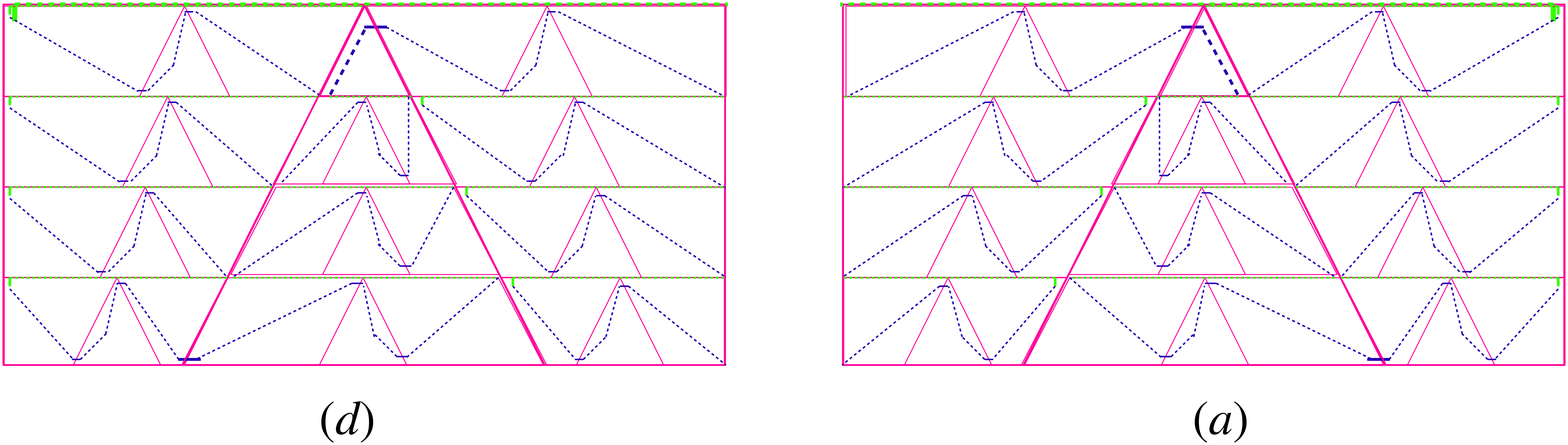,width=350pt}}
\vskip -15pt
\ligne{\hfill
\PlacerEn {-350pt} {-40pt} \box110
}

\vspace{-20pt}
\begin{fig}\label{quadrangle_deux}
\leurre
The splitting of the slices: second generation.
\vskip 0pt\parindent 0pt
On the left-hand side: the descending case. On the right-hand side:
the ascending case.
\end{fig}

   Now, these slices can again be split into four horizontal 
{\bf slices} defined by the four latitudes of mauve triangles of the 
just previous generation which are contained in the latitude of the 
slice which we consider. Figure~\ref{quadrangle_deux} illustrates the 
result of this splitting for a slice.

   The third point is that the slices almost follow the same
pattern, possibly after reflection in a vertical axis. We simply notice that
on one side, the lowest slice gives access to the triangle instead of
putting the path on the next isocline. On the other side, the topmost slice
also gives access to the triangle instead of going to the next isocline. 
Also, we have to note that the topmost region inside a triangle, see
figure~\ref{quadrangle_deux} is never a triangle. The representation
of the figure is due to the distortion introduced by the Euclidean
representation. The upmost slice inside a triangle is again a slice, even
if the part outside the triangle seems to be much smaller.
At last, we also notice that the green line of a slice is also the
green line of the topmost slice of the previous generation which is
just below the isocline of the line. As we assume that all mauve triangles
are finite, such a line will always meet a leg of a triangle, not at a
corner of the leg.

   Also, a last point to notice is that inside a slice there are several
triangles of the same generation within the latitude of the slice. In the
figures, we represented a single one due to the Euclidean constraints. But
in the hyperbolic plane there are a lot of them. Hower, from the figure
itself, it is not difficult to see that the same pattern is followed inside
the region delimited by two consecutive triangles: it is simply a trapeze.

   At last, note that we have to determine the isoclines which delimit
the different slices. The latitude of the slice is that of the mauve triangles
of a certain generation. We proceed as follows. The first slice 
goes from the vertex of the mauve triangle to the mid-point of its 
supporting red triangle~$R$. The second slice goes from this mid-point 
to the basis of~$R$: again something which is easy to determine on the 
mauve leg. The third and the last slices are clearly determined by the
low point and the corners of the mauve triangle. In sub-section~3.3 we have 
seen how to determine the low point on the leg of a mauve triangle.

   It is not difficult to see that in this way, provided that the path meets
legs of bigger triangles, which is the case from lemma~\ref{cover} and from
our assumption that there is no infinite mauve triangle, it will 
go further on the latitudes which it already visited and that it enters new 
latitudes. Thus, we can define times~$t_k$ satisfying the above conditions. 
Accordingly, the path fills up the plane and there is no initial tile.

   Now, we have to look at the schema much closer as mauve triangles
overlap, which leads us to precisely define the slices. We turn to this
point in the next sub-section.

\subsection{The tuning of the slices}
  
   If we look at the yellow frames of figure~\ref{lesmauves}, we can see
that the border of a triangle can be altered in different ways. First,
its basis is crossed by many mauve triangles of smaller generations. Next, it
is also possibly crossed by a mauve triangle of a bigger generation.

   Assume that the mauve triangle~$T$ we consider is not crossed by a mauve
triangle of a bigger generation. Then, we define the new slices by
simply following the border of the mauve triangles which intersect
the basis of~$T$ and, recursively, for the basis of each one of such triangles.
This process terminates on a mauve-1 triangle, see figure~\ref{lesmauves}.
In between two triangles of the same generation as~$T$, the lower border of
a trapeze is also determined in a similar way: it recursively follows the 
lower part of mauve triangles crossing the isocline determined by the corner
of the mauve triangle. Also, in this case, the lower exit/entry of~$T$
is at its left-hand side corner.

   Now, consider a mauve triangle~$T_1$ which is of the generation~$n$,
where $n$+1 is the number of the generation of~$T$, and assume that the legs 
of~$T_1$ cut the basis of~$T$. In~$T_1$, this intersection occurs along
the isocline which passes through the low point of its legs. This isocline
determines the upper border of the lowest slice of~$T$ up to the recursive
detours caused by mauve triangles of smaller generations. Now,
as the basis of $T_1$ is below the basis of~$T$, placing the entry to~$T_1$
at its left-hand side corner will force the path to cut itself. And so,
to avoid this point, we place the entry of~$T_1$ at the intersection of
the left-hand side leg of~$T_1$ with the basis of~$T$: it is the left-hand side
low point of~$T_1$. Now, it is not difficult to adapt the schemes of the
figure~\ref{tri_trap}, \ref{quadrangle} and~\ref{quadrangle_deux} 
to the new situation. We repeat this new definition of the entry each time
when a basis of a mauve triangle crosses the low point. Now, it is easy to
determine whether the entry of a triangle is at its left-hand side low point 
or at its corner. The signal of the entry will be present at the low point
if a basis is present. If a basis is not present, the low point sends a
signal to its corner, below, in order to trigger the signal of the entry.

   Now, it is easy to see that the same new definition of the lower entry
applies when $T$~is crossed by the basis of a bigger generation: it is
namely crossed at its low points and so the left-hand side one becomes
the entry of~$T$. The lower border of the slice associated to~$T$ is the
basis of the mauve triangle of a bigger generation which cuts~$T$. Also
note that, from the lower border of the slice which is just above~$T$, we
notice that the end of~$T$ around its vertex is cut by possibly a mauve
triangle of a smaller generation, involving the same kind of 'embroidery'
as in the lower border of the slice of~$T$. 

   As a last point, we have to indicate that the vertex of a mauve triangle
can no more be use for the exit from the triangle, unless it falls within a
slice of the generation~0. In the other cases, the exit of triangle~$T$
of positive generation is determined by the lower border of the upper slice
which cuts its legs. The exit occurs at the point which is on the leg, just 
below the border. The border is determined by the basis of a mauve triangle
of the previous generation which cuts~$T$.

   As the lower border of a slice is also the upper border of the next slice,
below, we completed the change which we had to introduce in order to
take into account the overlappings between mauve triangles.

   With these modifications, we can see that the travel of the path is
globally the same as what we described in the previous sub-section.

\subsection{About the tiles}   

   To conclude the proof of theorem~\ref{plane_fillka}, we have to 
give a few details about the implementation of the just described algorithm
in a finite set of tiles.

   We remember that the tiles are heptagons on which various signals run, 
defining the colours of the edges of the tiles.

   We have already all the signals inherited by the mantilla and by the
construction of the interwoven triangles. To this, we append the
signals for the construction of the mauve triangles and the determination 
of their low points.

   The last point is to describe how the path finds its way
from which the tiles can easily be deduced.

   For this purpose, we remember that most often, the path performs
zig-zags in between vertical borders, call them {\bf walls}. In between
two walls, the path runs along an isocline. It runs in one direction on one
isocline and in the opposite direction on the next isocline. We can have the 
same colour for both directions. However, we shall have tow colours: one for
an ascending path and the other for a descending one. Note that in different
realizations of the tiling, the colours may be interpreted in the opposite
way: what is important is that we have two colours. 

   Also, to facilitate the implementation, we have to prevent going from one 
isocline to another, when the path is not at a wall. The wall is always 
materialized by a mauve leg. Now, the path meets the wall on both sides: 
when it is inside the corresponding mauve triangle and when it is outside. 
We decide that the legs of the mauve triangles are always reached by a path 
which is inside the triangle. As the leg must stop outside portions of the 
path going to it, in fact the tiles of a mauve leg have a mark on the side
of the tile which is in contact with the outside of the triangle and which
is on the isocline. Remember that in \cite{mmarXivb,mmarXive}, we assign a 
local numbering to the sides of a tile. We number the side shared with the 
father by~1 and the other sides are increasingly numbered when counter-clockwise
turning around the tile. As any tile has a father, this fixes the local 
numbering of the tiles everywhere. Accordingly, on the left-hand side border,
the mark for outside parts of the path is on the side~3 of a tile. On the 
right-hand side border, the mark is on the side~7.

   We also have to note that the exact definition of the slices entail
a distortion of the path. What is represented by horizontal lines in
figures~\ref{quadrangle} and~\ref{quadrangle_deux} is not always along an
isocline. A few portions of the path go along a wall, stopped by the leg
and we have to take this into account: this does not raise big difficulties,
especially when we are outside the leg, but we have to be careful when we are 
inside. This point is masked by the Euclidean representation. We have to 
remember that inside a triangle, the number of tiles on an isocline from
one leg to the other is divided by at least~2 when we go from an isocline
to the previous one, in their numbering. There is a shrinking of the space
which entails distortions. This requires to propagate the information of the
presence of the wall as long as it is needed. This is not very difficult to
realize: we have to take into account that, on each border, if the path goes to
one tile from the leg on the isocline~$i$, it cannot go closer than the 
fourth tile from the leg on the isocline~$i$$-$1. The required distortion,
in order that the path visits all tile is not difficult to realize: the
3 tiles left on the isocline~$i$$-$1 have to be visited by the path on the
isocline~$i$, which is easy to realize. It is important to indicate that
only one portion of the path goes in this 'parallel' way which follows a leg:
the other parts are performed by zig-zags to which we apply the just
indicated constraint, as long as the zig-zag do not again meet the leg, 
directly.

   At last, we notice that the number of isoclines from a slice to another
is always even, as this is the case for mauve-1 triangles. From this,
there is no problem to apply the following scheme: the leaving part is
always on the right isocline. 

   All the marks needed by the previous indications are easy to implement 
and require a finite number of tiles only. As we know, 
from~\cite{mmnewtechund}, the number of tiles is huge, already for the 
construction of the interwoven triangles. 

   This completes the proof of theorem~\ref{plane_fillka}.

\subsection{The case of infinite triangles}

   As indicated in our introduction, from what we proved in the previous 
sub-sections shows that each time we have a tiling which does not generate
infinite mauve triangles, we get a uniform plane-filling path defined by 
the tiling.

   But this is no more true if there is an infinite mauve triangle. Once
the path falls inside such a triangle, it is trapped: later, it can never  
go outside the triangle.

   Now, such a situation is possible. From~\cite{mmarXivb,mmarXive}, we know
that in the case of a realization of the butterfly model, there are four
possible cases for the line~0 of the model: it may accompanied by a blue
basis, which brings no harm to our construction. It may be accompanied by
a red basis. If its a red basis of a triangle, it is an infinite red triangle,
but the corresponding infinite mauve triangle is removed to infinity: it
has no trace in the hyperbolic plane and so, this situation s also handled
by our construction. If the line~0 is accompanied by a basis of a red
phantom, this basis gives rise to red infinite triangles and, 
consequently, the red vertices of these triangles generate mauve triangles 
which are also infinite. And so, in this last case, we have infinitely many
infinite mauve triangles. 

   Accordingly, the path is broken into infinitely many components. However,
each component is a fully filling path of the region which is
delimited by the component. Also, the infinite red basis of a phantom is
also the basis of an infinite mauve triangle. Accordingly, as this basis 
never meet the leg of a mauve triangle, except at its corner, the green
part of the path always runs on this isocline without any possibility to
leave it. Above this basis, our construction provides us with a single
path which visits all tiles exactly once. Denote by~$\pi_1$ the path which 
runs along the infinite mauve basis~$\beta$ and by~$\pi_2$ the path which
visits all tiles above~$\beta$ exactly once. Then, we can say that $\pi_1$
joins~$\pi_2$ at infinity. Similarly, the paths defined by each infinite 
mauve triangle and the corresponding infinite trapezes join each other 
at infinity. We have that all these paths are pairwise disjoint and that
any tile of the plane is visited by exactly one of them and once.
Moreover, none of the paths is a cycle.

\section{A cellular automaton to implement a uniform plane-filling path}

   Now, we can prove that there is a cellular automaton which implements
a uniform plane-filling path.
   
   The proof is simple: it is enough to construct the automaton in such a
way that it controls the construction of the interwoven triangles in
such a way that there is no infinite red triangle. It is not difficult
to see that there are infinite mauve triangles if and only if there are
infinite red ones. Now, to avoid infinite red triangles, it is enough to
avoid a realization of the butterfly model.

   The idea for that is the following.

   The automaton will operate on two layers: we can see each tile as 
belonging to two copies of the ternary heptagrid. On one layer, the automaton
realizes the tiling and, on the second one, it controls the construction
performed on the first layer. Also, the automaton will draw the mantilla
and the numbering of the isoclines at the maximal speed, {\it i.e.} speed~1. 
The construction of the interwoven triangles is performed at speed at 
most~1, but there are longer and longer delays so that the construction
of the mantilla and the numbering of the isoclines are always in advance.
The construction of the mauve triangles and the path will be still slower.

   On the second layer, the automaton draws larger and larger circles around
a central cell~$\tau$, where by {\bf circle}, we mean the border of a ball
around~$\tau$. In fact, once a new circle is drawn, the old one is erased.
The r\^ole of the circle is to detect an isocline~15 which is not inside
a triangle. Such an isocline will be called {\bf void}. When the 
isocline~15 falls within a triangle, it is called {\bf covered}.

   The cellular automaton will start from a finite configuration: we take
a cell which will be, by definition, the place of the first active seed of 
an isocline~0. This also means that the cell is on the isocline~0. Then
the automaton constructs two blue triangles and the corresponding phantoms 
in between them. With the isocline~0 and the active seed, a line is defined by 
the cellular automaton: it is defined by the mid-points of the sides 2~and~6
of the active seed, while the isocline~0 crosses its sides~3 and~1. This is a
kind of vertical which will be used during the construction, it will
be called the {\bf initial vertical}. Then, 
the cellular automaton proceeds to the construction of the mantilla and the 
interwoven triangles until the first void isocline which occurs very soon:
an isocline~15 lies between the basis of the first blue-0 triangle and
the second ones which can be constructed a bit further on the next 
isocline~0. At this time, a cell~$\tau$ is chosen on the isocline~15 which
is defined by the intersection with a vertical issued from the active seed 
of the first seed, see \cite{mmarXivb,mmarXive} for the definition of such 
verticals.

   Now, it is decided that red triangles will be generated by the first
blue-0 triangle. In this way, the isocline~15 which passes through~$\tau$
is covered. From this time, the automaton grows a circle around~$\tau$
on the second layer, until the circle meets the first void isoclines:
these isoclines are detected by the fact that nothing exists on them
besides the numbering. When it is first realized, the circle meets 
two such isoclines as it works symmetrically with respect to the 
isocline~15 passing through~$\tau$. 

   First, the automaton constructs the generations~0 and~1 until it
si possible to define a triangle or a phantom of generation~2. Then, 
in order to cover the void isocline met by the circle above~$\tau$,
it is needed that a basis of a blue triangle is generated by the red triangle
previously determined. Similarly, to cover the other void isocline, the just 
generated triangle of the generation~2 generates a red triangle, accordingly
of the generation~3.

   Later, we proceed in this way:

   As soon as all void isoclines met by the circle are covered, the
construction stops and the circle is grown until a new void isocline appears,
below and above~$\tau$. The detection proceeds in this way: the circle 
advances by~20 steps, the distance between two consecutive isoclines~15. 
Each concerned cell of the circle sends a message to~$\tau$. If all isoclines 
are covered, $\tau$ sends a new message to go on by one move by~20 steps. 
This is repeated until $\tau$~receives the message that a void isocline is 
found in the appropriate direction: at each cycle of the construction, this 
direction is changed. It is initially upwards and then, it alternates. 

   When the void isocline is met, the growth of the circle is stopped.

   The construction of the interwoven circles is resumed. Note that, during
this time, the construction of the mantilla and the numbering of the isoclines
never stopped and during the phase which we shall soon describe, it also
never stops.

   The construction of the interwoven triangles is performed on all 
generations already constructed until it is possible to construct the 
first triangles of the next generation which will also cover the presently 
void isocline. The void isoclines are characterized by the fact that they 
have only the isocline number, here~15. The basis which accompanies a 
possible green line is not yet determined. Now, the definition of the next 
generation is made possible by the construction of at least two consecutive 
triangles of the same generation and along the same initial vertical. 
Then, the automaton can easily decide whether the needed triangle has its
vertex or its basis in the just previously constructed triangle.
Indeed, it is plain that the mid-distance line between the just determined
triangle and the next one along the initial vertical and outside the circle
is void. There may be void lines before as well but, in any case, the
triangle of the next generation which is built on these triangles
covers this mid-distance line and, a fortiori any one which would be closer 
to~$\tau$. The alternation of the direction guarantees that the construction
will cover the whole hyperbolic plane.

   Once the required triangle is constructed, the automaton goes on 
constructing all triangles of the previous generations which fall withing the ball delimited by the circle. Also, at this time, the corresponding mauve
triangles are constructed as well as the path. 

   When this is done, the construction is stopped and a new cycle is 
performed: first by growing the circle again, and then, by performing the 
required constructions.

\ifnum 1=0 {
the automaton draw the circle and make it growing
until it reaches the first void isocline. Also, it defines a 'vertical' as a 
line which is drawn along the mid-point of the sides 2~and~6, while the
isocline~0 crosses the sides~3 and~1. Then, it proceeds to the construction
of the mantilla and the interwoven triangles until the first void isocline
below~$\tau$ is met. This meeting fixes the blue-0 triangle rooted at~$\tau$
as generating a red triangle. As this triangle will cross the void isocline
detected by the circle, we continue the construction of this triangle and the
corresponding phantoms. The automaton then stops the construction and
it checks whether all isoclines~15 it crosses are not void. If still one
is void, which is the case at the beginning, the construction of the current 
generation is continued until a new triangle can be constructed, covering the
void isocline. Then, the automaton pushes the circle to the next void 
isocline, this time above~$\tau$. Then the construction of the interwoven 
triangles is resumed. The automaton constructs the tiling until it covers 
the ball around~$\tau$ delimited by the circle. During this construction, 
each time a vertex has to be placed, it is checked whether the choice which 
is made will cover the current void isocline. There is always a choice 
which satisfies this requirement.

   The construction goes on this way, ruling the growth of the circle by the 
meeting of the first void isocline starting from~$\tau$, alternatively above 
and below~$\tau$. This allows to guarantee that any triangle is contained in
some red triangle of a bigger generation or in between two such consecutive
triangles within the same latitude. Also, the growing of the circle makes
it certain that the construction will provide a tiling of the plane. The
mauve triangles are drawn at the same time as the interwoven triangles which
allow to draw the path starting from the first vertex of a red triangle. For
the automaton, drawing a signal on a tile simply means changing this tile
with a new one where the new signal is appended to those already indicated
in the tile. This means, in particular that the first time isoclines~15 are 
detected, the corresponding tiles only bear the mark of an isocline~15.

   The detection of the void isocline and the process of enlarging the
circle can be efficiently organized by a synchronization mechanism. When
a vertex is placed which guarantees that the void isocline will be covered,
then, a signal goes from the 'void' on the circle to~$\tau$. When it 
reaches~$\tau$, the central cell sends back a synchronizing signal to
the circle at speed~1. It is clear that all cells of the circle are reached 
exactly at the same time. The increasing of the radius is known in advance:
it is by~20 cells: then the next isoclines~15 are met 
}
\fi

   Accordingly, as the automaton covers larger and larger balls around the
same tile~$\tau$, it constructs the path in infinite time. Also, as
the process guarantees that the void isoclines are step by step covered,
there is no infinite triangle. And so, the construction of a path
satisfying the assumptions of theorem~\ref{plane_fillka} is achieved in
infinite time, which proves theorem~\ref{CA_fillka}.

\section{A Peano curve}

   Now, a Peano curve can easily be constructed. The topic is not 
new and has been delt with in much larger contexts than this one, 
see~\cite{thurston}. However, we can give a constructive implementation
based on the uniform plane-filling path which is obtained 
in section~5.

   The path fixes an order on the tiles so that we can number the tiles 
with~$Z\!\!\!Z$, assigning~0 to an arbitrary tile, fixed once for all.
For the step~0, we proceed as follows. In each tile, two of its edges, say~$i$
and~$j$ are marked as ends of the path. consider the mid-points $A$ and $B$
of the edges~$i$ and~$j$ respectively. The points~$A$ and~$B$ determine a
segment which is supported by the hyperbolic line passing through~$A$ and~$B$.
In the considered tile~$n$, the trace of path in~$n$ is the segment~$[AB]$.
  
\setbox110=\hbox{\epsfig{file=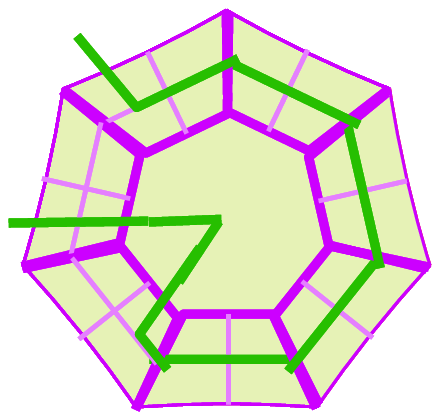,width=200pt}}
\setbox112=\hbox{\epsfig{file=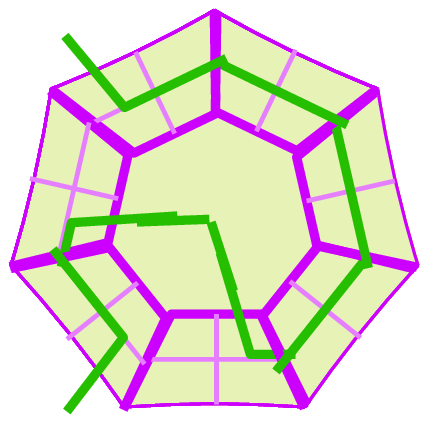,width=200pt}}
\setbox114=\hbox{\epsfig{file=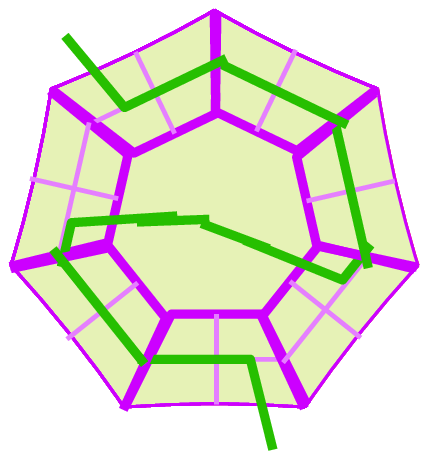,width=200pt}}
\setbox116=\hbox{\epsfig{file=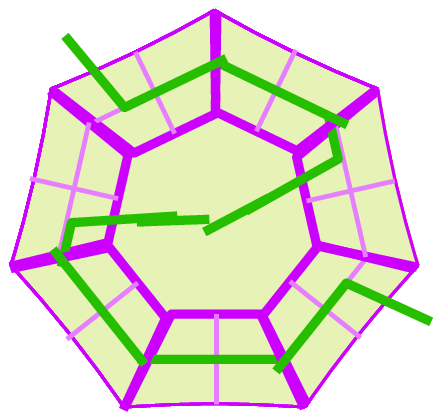,width=200pt}}
\setbox118=\hbox{\epsfig{file=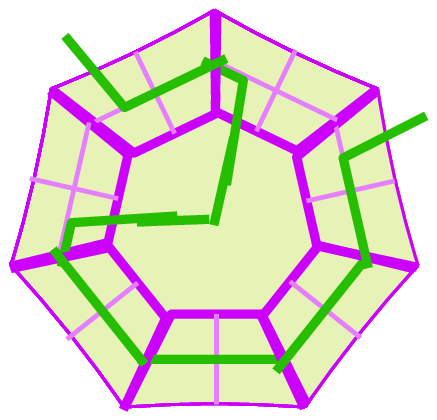,width=200pt}}
\setbox120=\hbox{\epsfig{file=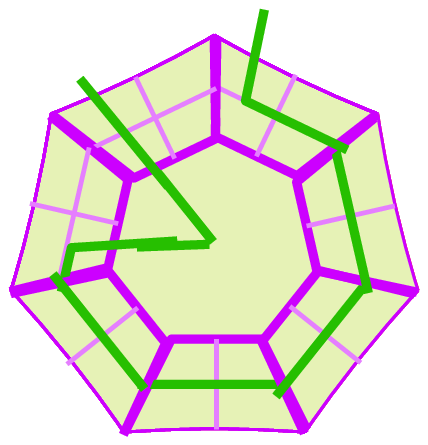,width=200pt}}
\vskip -45pt
\vtop{
\ligne{\hfill
\PlacerEn {-380pt} {0pt} \box110
\PlacerEn {-280pt} {0pt} \box112
\PlacerEn {-180pt} {0pt} \box114
}
\vskip -110pt
\ligne{\hfill
\PlacerEn {-380pt} {0pt} \box116
\PlacerEn {-280pt} {0pt} \box118
\PlacerEn {-180pt} {0pt} \box120
}
\vspace{-50pt}
\begin{fig}\label{peano_basis}
\leurre
The first step of the downwards construction.
\end{fig}
}
\vskip 5pt

   For the next step, in each tile~$n$, we replace the segment of the path 
which crosses~$n$ by the appropriate path represented in 
figure~\ref{peano_basis}. The tiles of the figure represent all the possible 
cases for an entry on a fixed edge and the exit on another one: there are 
indeed six~possibilities. These basic patterns can be adapted by an 
appropriate rotation around the centre of the tile so that one of the
entries coincide with the side of the tile crossed by the path of step~0.
This defines the path of step~1.

   Note that each pattern of figure~\ref{peano_basis} can be split into
two parts: the {\bf central} one and the {\bf ring}~1. The ring~1 can be viewed
as splitted into seven parts. Such a part is a trapeze, defined by an edge of 
the tile, two radiuses of the tile, from its centre to the vertices of this 
edge, and the segment joining the mid-points of the radiuses. The mid-points 
of the radiuses from the centre~$\Omega$ of the tile and its vertices, are 
called the points of {\bf order}~1.

   Now, inside a tile~$n$, the points of order~1 define a new tile which we 
denote by~$n_1$. Note that the centre of~$n_1$ is also~$\Omega$. Now, the 
tile~$n_1$ is crossed by a path defined by two rays joining at~$\Omega$, 
the centre of the tile. Note that~$n_1$ is also a regular heptagon, but its
angles are not those of~$n$ as it is smaller. Remember than in the hyperbolic
plane there is no similarity.

\setbox110=\hbox{\epsfig{file=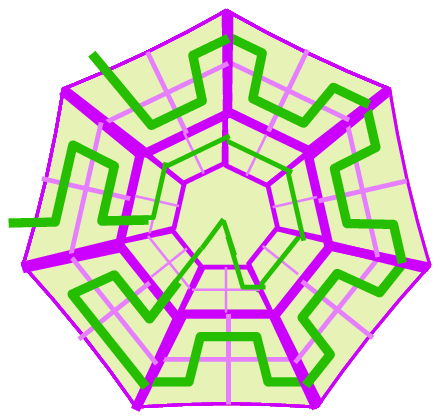,width=200pt}}
\setbox112=\hbox{\epsfig{file=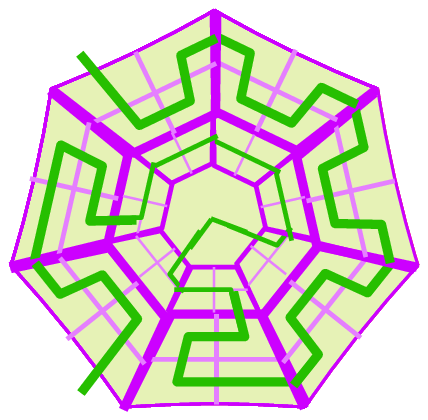,width=200pt}}
\setbox114=\hbox{\epsfig{file=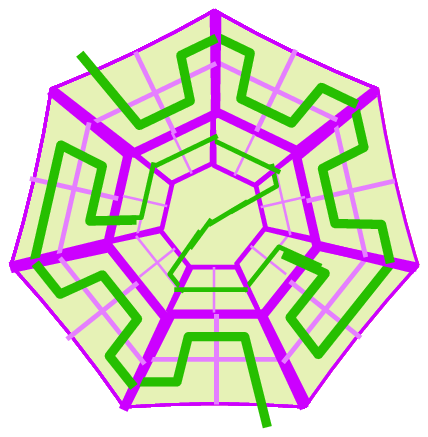,width=200pt}}
\setbox116=\hbox{\epsfig{file=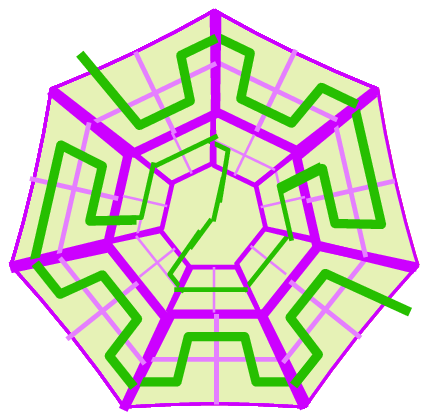,width=200pt}}
\setbox118=\hbox{\epsfig{file=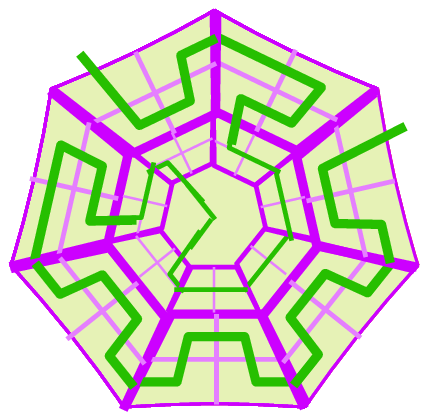,width=200pt}}
\setbox120=\hbox{\epsfig{file=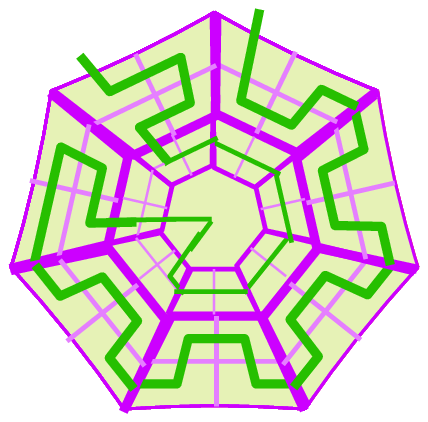,width=200pt}}
\vskip -45pt
\vtop{
\ligne{\hfill
\PlacerEn {-380pt} {0pt} \box110
\PlacerEn {-280pt} {0pt} \box112
\PlacerEn {-180pt} {0pt} \box114
}
\vskip -110pt
\ligne{\hfill
\PlacerEn {-380pt} {0pt} \box116
\PlacerEn {-280pt} {0pt} \box118
\PlacerEn {-180pt} {0pt} \box120
}
\vspace{-50pt}
\begin{fig}\label{peano_next}
\leurre
The second step of the downwards construction.
\end{fig}
}
\vskip 5pt

   Assume that we defined the step~$k$. In each tile~$n$, we define 
in~$n_1$ the same construction as we defined for~$n$ at the step~$k$.
If we were in the Euclidean plane, we could simply say that we apply to
$n_k$ a dilatation around~$\Omega$ of amplitude~$\displaystyle{1\over2}$.
Now, in the hyperbolic plane such dilatations do not exist. However, we can
repeat the construction which we defined for~$n$ to~$n_1$, as it is based on 
a two dimensional dichotomic process. Now, we have to also define how we
transform the ring~1. In each trapeze of the ring, we have cells of order~$k$
which are also trapezes. We can see this when we go from the step~1 to 
the step~2, see figures~\ref{peano_basis} and~\ref{peano_next}. Each trapeze
of order~$k$ has opposite sides and opposite bases, defined at this stage.
When $k=1$, the sides are supported by the radiuses which define the trapezes
and the bases are an edge of~$n$ and the corresponding edge or~$n_1$. Each
trapeze of order~$k$ is crossed by a segment from one side to the other or 
a segment crosses one of the bases and then crosses a side. For the 
step~$k$+1, each trapeze is split into four trapezes of order~$k$+1: we join 
the mid-points of the sides and the mid-points of the bases and this define 
the four new trapezes. In the new trapezes, what is on a side of the trapeze 
of order~$k$ remains a side and what is a basis remains a basis: this allows 
to define the sides and the bases of the trapezes of order~$k$+1. An edge
of a trapeze of order~$k$+1 is a side of order~$k$+1 if and only if either 
it is supported by a side of order~$k$, or if it is opposite to a side of 
order~$k$. Similary, an edge of a trapeze of order~$k$+1 is a basis of 
order~$k$+1 if and only if either it is supported by a basis of order~$k$
or it is opposite to a basis of order~$k$.

   In fact the pattern of the path inside a trapeze can be described in
a precise way which is very close to what is performed in the Euclidean
plane, for instance in the construction of the plane-filling path 
of~\cite{jkari94}. However, in the quoted paper, the construction defines
growing up structures and here, we go in the opposite direction: the new
structures are smaller and smaller. Figure~\ref{peano_split} indicates 
the patterns which are used to go from the step~$k$ to the step~$k$+1{}
in the ring~1. 

\setbox110=\hbox{\epsfig{file=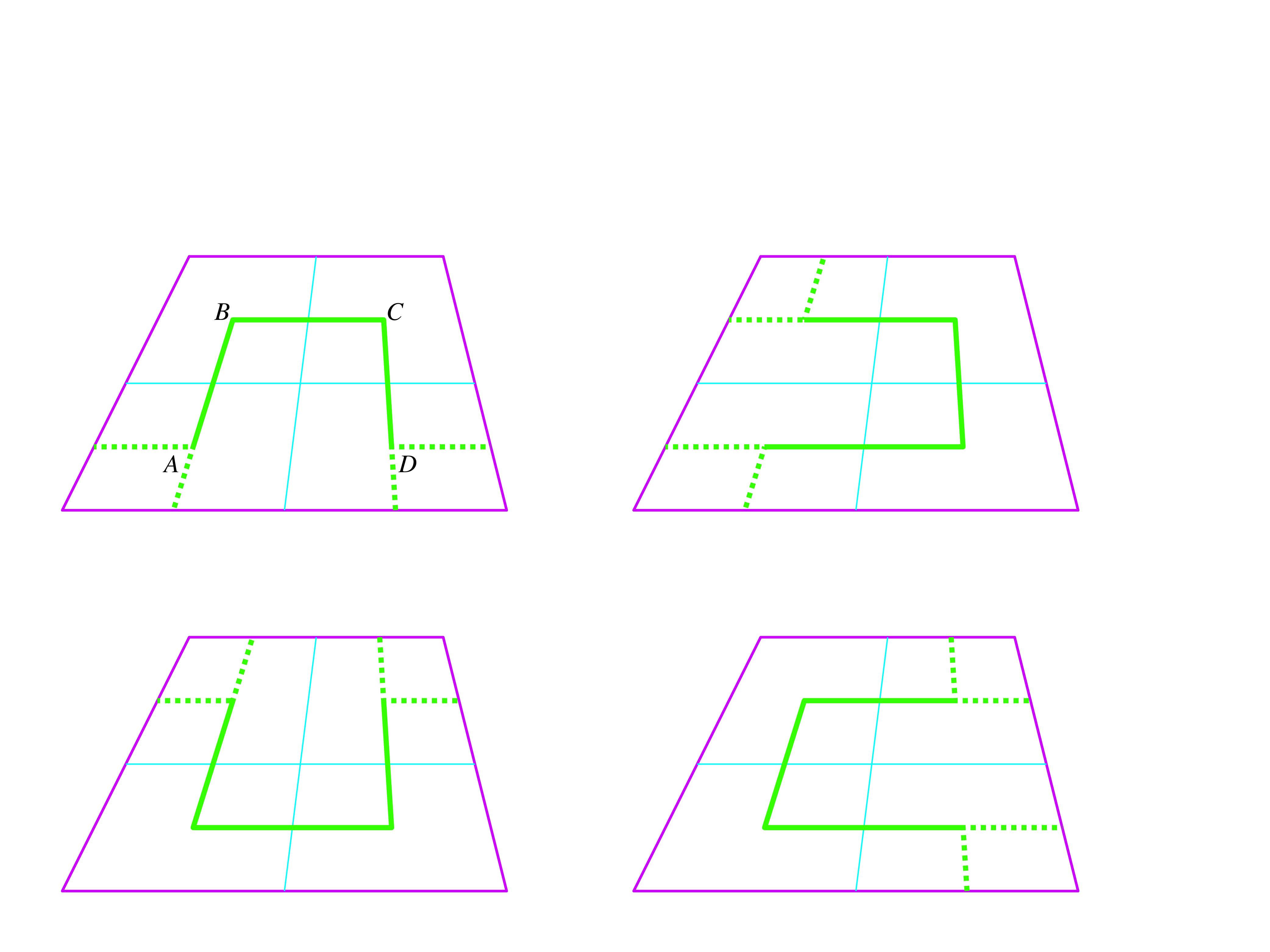,width=350pt}}
\vskip -50pt
\vtop{
\ligne{\hfill
\PlacerEn {-350pt} {0pt} \box110
}
\vskip -130pt
\begin{fig}\label{peano_split}
\leurre
From the step~$k$ to the step~$k$$+$$1$.
\end{fig}
}
\vskip 5pt
   The four trapezes of the figure indicates the four possible paths
which can be symbolized by $ABCD$, $BCDA$, $CDBA$ and~$DABC$.
Now, the connection with a neighbouring trapeze or with the central region   
is given by paths which are represented by the dotted lines of
figure~\ref{peano_split}. In each picture of the figure, only two points
among~$A$, $B$, $C$, $D$ give rise to dotted lines. These points are
called the {\bf entries}. Now, at each entry, the path goes through one dotted
line exactly. And so, each picture gives rise to four paths, depending on the 
connection with the neighbouring ones.

   As $k$~becomes bigger, the situation looks closer and closer to a Euclidean 
situation. Indeed, as $k$~gets bigger, the size of the trapezes of order~$k$ 
becomes smaller and smaller. Now, it is known that the infinitesimal elements 
are the same for the Euclidean and for the hyperbolic planes. This means that,
as the neighbourhoods of a point get smaller, the hyperbolic situation looks
more and more Euclidean.

   Now, for each~$k$, the step~$k$ defines an infinite curve~$P_k$. We also 
know that, locally, the Euclidean plane and the hyperbolic one have the same 
topology. Consequently, we can see that the curves~$P_k$ simply converge to
a curve~$P_\infty$ which goes through each point of the hyperbolic plane.

   As indicated in the introduction, it is possible to define a simpler
construction of a Peano curve than this one: it will be done in a 
forthcoming paper. It also makes use of the figures~\ref{peano_basis},
\ref{peano_next} and~\ref{peano_split} of this section. 

\section{Conclusion}

   I hope that this paper shows the interest of the construction given
in \cite{mmarXivb,mmDavidson,mmnewtechund}. As indicated in
\cite{mmarXiv4}, there is still much work to do in this domain. In particular,
it remains to see whether the strong plane-filling property holds or not 
in the hyperbolic plane.




\end{document}